\documentclass[12pt,a4paper]{article}

\usepackage{amsmath}
\usepackage{amssymb}
\usepackage{epsfig}
\usepackage{graphics}

\def\nostrocostrutto#1\over#2{\mathrel{\mathop{\kern 0pt \rlap 
  {\raise.2ex\hbox{$#1$}}}
  \lower.9ex\hbox{\kern-.190em $#2$}}}

\catcode`@=11
\newcount\@tempcntc
\def\@citex[#1]#2{\if@filesw\immediate\write\@auxout{\string\citation{#2}}\fi
  \@tempcnta\z@\@tempcntb\m@ne\def\@citea{}\@cite{\@for\@citeb:=#2\do
    {\@ifundefined
       {b@\@citeb}{\@citeo\@tempcntb\m@ne\@citea\def\@citea{,}{\bf ?}\@warning
       {Citation `\@citeb' on page \thepage \space undefined}}%
    {\setbox\z@\hbox{\global\@tempcntc0\csname b@\@citeb\endcsname\relax}%
     \ifnum\@tempcntc=\z@ \@citeo\@tempcntb\m@ne
       \@citea\def\@citea{,}\hbox{\csname b@\@citeb\endcsname}%
     \else
      \advance\@tempcntb\@ne
      \ifnum\@tempcntb=\@tempcntc
      \else\advance\@tempcntb\m@ne\@citeo
      \@tempcnta\@tempcntc\@tempcntb\@tempcntc\fi\fi}}\@citeo}{#1}}
\def\@citeo{\ifnum\@tempcnta>\@tempcntb\else\@citea\def\@citea{,}%
  \ifnum\@tempcnta=\@tempcntb\the\@tempcnta\else
   {\advance\@tempcnta\@ne\ifnum\@tempcnta=\@tempcntb \else \def\@citea{--}\fi
    \advance\@tempcnta\m@ne\the\@tempcnta\@citea\the\@tempcntb}\fi\fi}
\catcode`@=12

\begin{document}

\setcounter{page}{0}
\thispagestyle{empty}
\begin{titlepage}

\vspace*{-1cm}
\hfill \parbox{3.5cm}{BUHE-01-01 \\ 
31 March  2001
\vspace*{0.3cm}
 }   
\vfill

\begin{center}
  {\large {\bf
  The strange border of the QCD phases\footnote{We thank the Schweizerische Nationalfonds for his support.}  }
\vfill
\vspace*{0.3cm} 

{\bf
    Sonja Kabana } \\
    Laboratory for High Energy Physics \\
    University of Bern \\
    CH - 3012 Bern, Switzerland
    \\
    E-mail: sonja.kabana@cern.ch
   \vspace*{0.3cm} 
   \vspace*{0.3cm} 
}

\end{center}

\vfill

\begin{abstract}
\noindent
We address the flavour composition along the border between the
hadronic and the quark-gluon plasma phases of QCD.
The ratio of strange to up and down  antiquarks ($\lambda_s$) produced in particle and nuclear collisions,
is found to increase in collisions with initially reached energy density
($\epsilon_i$) up to $\epsilon_{crit}$ $\sim$ 1  GeV/$fm^3$.
Above this value it decreases approximately linearly
and reaches its asymptotic value
at zero baryon chemical potential ($\mu_B$).
We demonstrate that $\lambda_s$ in nuclear collisions
is approaching its asymptotic value 
at $\epsilon_i$ $\sim$ 8-9 GeV/$fm^3$, corresponding
 to $\sqrt{s}$ $\sim$ 3-8 TeV per nucleon+nucleon pair
which will be reached at the LHC.
After correcting for the difference in the chemical potentials 
of  various colliding systems,
$\lambda_s$ universally saturates  across the 
QCD phase boundary, following the temperature.
Recent experimental puzzles as the increase in the $K/\pi$ ratio
in Pb+Pb collisions at 40 GeV per nucleon, its different behaviour
at midrapidity, the decrease
of the double ratio of $K/\pi$(A+A/p+p) in nucleus nucleus over
p+p collisions with increasing $\sqrt{s}$, 
and the increase of $\lambda_s$ in p+A over p+p
collisions at the same $\sqrt{s}$, are naturally explained.
We study
the approach of thermodynamic observables at $\mu_B=0$
to the transition point and
extract an estimate of the critical temperature.

\end{abstract}

\vfill
\end{titlepage}

\newpage

\section{Introduction}

\noindent
One outstanding prediction of the theory of strong interaction
(e.g. \cite{lattice,GL,PM,mapping}),
which applies 
to the evolution of the early universe, is the phase transition from
confined hadrons to 
a deconfined phase of their constituents, the
quarks and gluons (the so called quark gluon plasma =QGP).
An experimental program which started approximately in the 
eighties and continues
in many accelerators as: CERN SPS, BNL  RHIC and others,
has been dedicated to the experimental verification of this 
transition \cite{qmconference,stock,heinzjakob,satz,horst}.
The main process is nuclear reactions at high energy to achieve
a thermalized state with temperature exceeding the 
critical temperature for the QCD phase transition.
One of the predicted signatures of this change of phase, is
an enhancement of strange  particles \cite{rafelski1}.
The main idea was a)
the lower threshold for the production of
$s \overline{s}$ quarks 
in the QGP through e.g. $gg \rightarrow s \overline{s}$
as compared to the higher threshold for production of
strange hadrons in a hadronic reaction for example
$p p \rightarrow p \Lambda K^+$
and b) the similarity of the mass of the strange quark with the
critical temperature $T \sim 200 MeV$, which allows for the equilibration of
strange quarks in the QGP.
\\
Argument a) holds for all quark flavours, leading to the
conclusion that all hadron multiplicities are expected to be increased
when produced out of a hadronizing QGP, as compared to hadron production
out of a collision which does not pass through the QCD phase transition.
The enhancement can  be related in a simple way to phase space
\cite{Sonja} 
and is flavour dependent.
\\
However the argument b) is true only for
the quarks which have mass less or of the order of the critical
temperature of approximately 200 MeV \cite{mapping}.
\\
Therefore next to the  up and down quarks, strange
quarks are expected to 
play a crucial role in identifying the QCD phase transition.
In particular, hadrons with 
$u, \overline{u}, d, \overline{d}, s$ and $\overline{s}$ quarks are expected to
reflect 
the critical behaviour of  a locally equilibrated phase made up of
$u, \overline{u}, d, \overline{d}, s, \overline{s}$ quarks and gluons.
\\
Heavier flavours like charm and beauty 
 can be affected by the QCD transition
in several ways discussed in the literature 
\cite{jpsi_prediction,satz,horst_charm}.
They give important insights e.g. by
the experimental measurement of the dissociation temperature
of $c \overline{c}$ or $b \overline{b}$
states which can be at  or above the critical one
\cite{jpsi_prediction,satz}.
\\
Therefore, since dissociation of quarkonia may be due to overcritical
energy densities, it appears that the critical parameters of the transition
can be extracted only through other observables which are sensitive
to the transition at the critical energy density and
approximately frozen with fast hadronization. For example
fluctuations of several parameters and
 multiplicities
of hadrons which include up, down and strange quarks.
\\

\noindent
In this paper we concentrate on the behaviour of the strange
flavour across the QCD phase boundary.
The main point is to interpret the global features of the data 
(A+A, $p \overline{p}$, $e^+ e^-$ collisions
at $\sqrt{s}$=2-1800 GeV)
within a strictly thermodynamic
approximation, and not to reproduce the data in detail using
models outside the  validity of thermodynamics.
\\
Clearly, no perfect global equilibration is reached in these systems
which moreover are a mix of several thermodynamic reservoirs e.g.
along the rapidity axis.
It appears however that the hadrons in the final state of the colliding systems
studied, do not rule out a thermodynamic description.
This is  especially the case when a centrality selection has been
imposed.
\\
Preliminary results of part of this work  have been shown in \cite{moriond}.
\\

\noindent
In section 2 we extract thermodynamic parameters $(T, \mu_B, \mu_s)$ for
fixed target Au+Au collisions at 2 and 4 GeV per nucleon 
and for fixed target 
Pb+Pb collisions at 40 GeV.
In section 3 we estimate the initial energy density reached in
the collisions.
In section 4 we examine the 
strangeness suppression factor at non zero chemical potentials
as a function of the initial energy density,
and find its asymptotic value using nuclear collisions.
In section 5 we extract the temperature at zero chemical potentials
and draw the 'strange border of the QCD phases'.
We discuss how the critical parameters for the transition are
extracted and elaborate on the explanation of several
recent experimental puzzles.
\\

\section{Thermodynamic description of nuclear collisions at 2, 4 and 40 A GeV}

\noindent
We compare the ratios of experimentally measured
hadron yields in nuclear collisions with the prediction of
a grand canonical ensemble of non interacting free hadron resonances.
We consider
 the pseudoscalar and vector u,d,s meson nonets as well
as the spin 1/2 baryon octet and spin 3/2 decuplet and their
antiparticles as well as the $f_0(400-1200)$ or $\sigma$, interpreted
as scalar glueball \cite{SKPM}.
We impose exact conservation of strangeness and correct
for the change of final observed hadron yields due to decays
of higher lying resonances (the so called 'feeding').
Further details of this model can be found in \cite{mapping}.
We extract then
the thermodynamic parameters:  temperature, baryochemical potential
($\mu_B$)
and strangeness chemical potential ($\mu_S$)
from the model prediction which descibes the data best. 
We discuss in the following the production of strangeness relativ
to non strange particles extracting  the quantity:
$
\lambda_s \ = \
\frac{ (2 \overline{s}) }
{ (\overline{u} + \overline{d}) }
$
which is a measure of the strangeness suppression factor
defined and used in the literature as
$
\lambda_s \ = \
\frac{ 2 (s + \overline{s}) }
{ (u + \overline{u} + d + \overline{d}) }
$
\cite{revions1}.
\\
Similar analysis extracting thermal parameters from data,
can be found in the literature e.g.
\cite{hagedorn,Gerber,revions,revions1,rafelski,biro,becattee,hepph9702274}.
For a recent review see \cite{rischke}.
\\
A main new idea  introduced in 
\cite{Sonja,mapping} and used also in this paper as a tool,
is the extrapolation of all thermodynamic states to equivalent states
at zero chemical potentials e.g. along isentropic paths.
\\

\noindent
Throughout the paper, all parameters discussed 
concern the state of hadrons at their chemical freeze-out, that is
the time after which the hadron yields do not change anymore
through inelastic collisions.
The only exception is the initial energy density which 
refers to the early state of the system.

\subsection{Thermodynamic description of Au+Au collisions at 2 A GeV}

\noindent
We use 3 ratios measured in fixed target Au+Au collisions at 2 GeV per nucleon
beam energy shown in table \ref{ratios_auau2gev}
from references \cite{pinkenburg,e917}
and impose strangeness conservation. 
The predicted 
and the experimental ratios are shown in table \ref{ratios_auau2gev}.
The $K/\pi$ and the $\pi/p$ ratios are measured at midrapidity,
while the $\Lambda / K^0_s$ ratio nearby \cite{pinkenburg}.
To account for the different phase space we add a 10\% systematic error to the
$\Lambda/ K^0_s$ ratio.
\\
The resulting $\chi^2/DOF$ is (1.41/1) (CL 23\%).
After defining the (T, $\mu_b$, $\mu_s$) values describing the particle
ratios produced in central Au+Au collisions at 
$\sqrt{s}$=2.3 GeV at the chemical freeze-out 
we extrapolate to the T at zero fugacities (table \ref{t_auau2gev})
along an isentropic path.

\begin{table}[ht]
\begin{tabular}{|lll|}
\hline
ratio  & model & data  \\
\hline

$K / \pi$  & 0.0110  & 0.00929 $\pm$ 0.00257 \\

$\pi / p$  & 0.268  & 0.266 $\pm$ 0.0770 \\

$ \Lambda/ K^0_s$  & 2.770  & 3.222 $\pm$ 2.274 \\

\hline
\end{tabular}
\caption{
Au+Au at $\sqrt{s}$= 2.3 GeV. \protect\newline
Predicted versus experimental particle ratios for the best fit of our model.
}
\label{ratios_auau2gev}
\end{table}

\begin{table}[ht]
\begin{tabular}{|lllllll|}
\hline
$\mu_b$   &  $\mu_s$     &  T   & $\lambda_s$ &  
$\rho_s$  &  $T({eq, \rho_s})$   & $\lambda_s({eq, \rho_s})$   \\
GeV       &  GeV         &  GeV &             &
1/fm$^3$ & GeV               &     \\
\hline

0.714  &   0.0918   &  0.0510  & 0.0321  & 0.0435   &   0.072  &  0.0388  \\

    &     &   $\pm$ 0.006 &  $\pm$ 0.0327  &    &  +0.032 -0.013 &  +0.116 -0.0249 \\

\hline
\end{tabular}
\caption{ 
Au+Au at $\sqrt{s}$= 2.3 GeV. \protect\newline
Thermodynamic parameters for the best fit and temperatures
and $\lambda_{\ s}$
extra\-polated to zero fugacities.
}
\label{t_auau2gev}
\end{table}

\subsection{Thermodynamic description of Au+Au collisions at 4 A GeV}

\noindent
We use 4 ratios measured in fixed target Au+Au collisions at 4 GeV per nucleon
beam energy shown in table \ref{ratios_auau4gev}
from references \cite{pinkenburg,e917}
and impose strangeness conservation. 
The predicted 
and the experimental ratios are shown in table \ref{ratios_auau4gev}.

The $K/\pi$, $\pi/p$ and $K^+/K^-$ ratios are measured at midrapidity,
while the $\Lambda/K^0_s$ ratio nearby \cite{pinkenburg}.
To account for the different phase space we add a 10\% systematic error to the
$\Lambda/K^0_s$ ratio.
\\
The resulting $\chi^2/DOF$ is (1.62/2) (CL $\sim$ 40\%).
After defining the (T, $\mu_b$, $\mu_s$) values describing the particle
ratios produced in central Au+Au collisions at 
$\sqrt{s}$=3.3 GeV at the chemical freeze-out 
we extrapolate to the T at zero fugacities (table \ref{t_auau4gev})
along an isentropic path.

\begin{table}[ht]
\begin{tabular}{|lll|}
\hline
ratio  & model & data  \\
\hline

$K / \pi$  &     0.0396    & 0.0422 $\pm$ 0.0111 \\

$K^+ / K^-$  &   12.437  & 12.316 $\pm$ 0.699 \\

$ \Lambda/ K^0_s$  & 2.342  &  2.750 $\pm$ 0.342 \\

$\pi / p$  &  0.473  &  0.435 $\pm$ 0.114 \\

\hline
\end{tabular}
\caption{
Au+Au at $\sqrt{s}$= 3.3 GeV. \protect\newline
Predicted versus experimental particle ratios for the best fit of our model.
}
\label{ratios_auau4gev}
\end{table}

\begin{table}[ht]
\begin{tabular}{|lllllll|}
\hline
$\mu_b$   &  $\mu_s$     &  T   & $\lambda_s$ &  
$\rho_s$  &  $T({eq, \rho_s})$   & $\lambda_s({eq, \rho_s})$   \\
GeV       &  GeV         &  GeV &             &
1/fm$^3$ & GeV               &     \\
\hline

0.657  &   0.0923   &  0.073  & 0.139  & 0.163   &   0.097  &  0.125  \\

    &     &   $\pm$ 0.005 &  +0.047 -0.040 &    &  +0.016 -0.009 &  +0.0710 -0.0359 \\

\hline
\end{tabular}
\caption{ 
Au+Au at $\sqrt{s}$= 3.3 GeV. \protect\newline
Thermodynamic parameters for the best fit and temperatures
and $\lambda_{\ s}$
extra\-polated to zero fugacities.
}
\label{t_auau4gev}
\end{table}

\subsection{Thermodynamic description of Pb+Pb collisions at 40 A GeV}

\noindent
We use 5 ratios measured in Pb+Pb collisions at 40 GeV per nucleon
beam energy shown in table \ref{ratios_pbpb40gev}
from references \cite{40gevpbpb,na49_kpi}
and impose strangeness conservation. 
The predicted 
and the experimental ratios are shown in table \ref{ratios_pbpb40gev}.
\\
The ratios $ \overline{\Lambda} / \Lambda $, $ \overline{\Xi} / \Xi $
are measured at midrapidity,
while the other ratios are given in full phase space acceptance.
The $(B - \overline{B}) $ is taken equal to the 
total number of participant nucleons.
\\
To account for the different phase space and the use of the 
assumption
$(B - \overline{B}) $= $N_{participant}$ we add a 10\% systematic error to the
$\pi/ (B - \overline{B}) $, $ \overline{\Lambda} / \Lambda $ and
$ \overline{\Xi} / \Xi $ ratios.
\\

The resulting $\chi^2/DOF$ is (6.68/3) (CL $\sim$ 10\%).
After defining the (T, $\mu_b$, $\mu_s$) values describing the particle
ratios produced in central Au+Au collisions at 
$\sqrt{s}$=8.76 GeV at the chemical freeze-out 
we extrapolate to the T at zero fugacities (table \ref{t_pbpb40gev})
along an isentropic path.

\begin{table}[ht]
\begin{tabular}{|lll|}
\hline
ratio  & model & data  \\
\hline

$K / \pi$  &     0.128    &  0.125 $\pm$ 0.0072 \\

$K^+ / K^-$  &   3.0355   &  3.163   $\pm$ 0.232 \\

$\pi / (B - \overline{B}) $  &   1.122  &  0.852  $\pm$ 0.125 \\

$ \overline{\Lambda} / \Lambda $  &  0.0216   &  0.0230  $\pm$ 0.00251 \\

$ \overline{\Xi} / \Xi $  &  5.146E-2 &   8.00E-2 $\pm$  2.625E-2 \\

\hline
\end{tabular}
\caption{
Pb+Pb at $\sqrt{s}$= 8.76 GeV. \protect\newline
Predicted versus experimental particle ratios for the best fit of our model.
}
\label{ratios_pbpb40gev}
\end{table}

\begin{table}[ht]
\begin{tabular}{|lllllll|}
\hline
$\mu_b$   &  $\mu_s$     &  T   & $\lambda_s$ &  
$\rho_s$  &  $T({eq, \rho_s})$   & $\lambda_s({eq, \rho_s})$   \\
GeV       &  GeV         &  GeV &             &
1/fm$^3$ & GeV               &     \\
\hline

0.405  &   0.090   &  0.150  &   0.660   &   2.466  &   0.164  &  0.407  \\

    &     &   +0.019 -0.039   &  +0.138 -0.326  &    &  +0.024 -0.049 &  +0.068 -0.199 \\

\hline
\end{tabular}
\caption{ 
Pb+Pb at $\sqrt{s}$= 8.76 GeV. \protect\newline
Thermodynamic parameters for the best fit and temperatures
and $\lambda_{\ s}$
extra\-polated to zero fugacities.
}
\label{t_pbpb40gev}
\end{table}

\begin{figure}[ht]
\vspace*{-0.5cm}
\begin{center}
\mbox{\epsfig{file=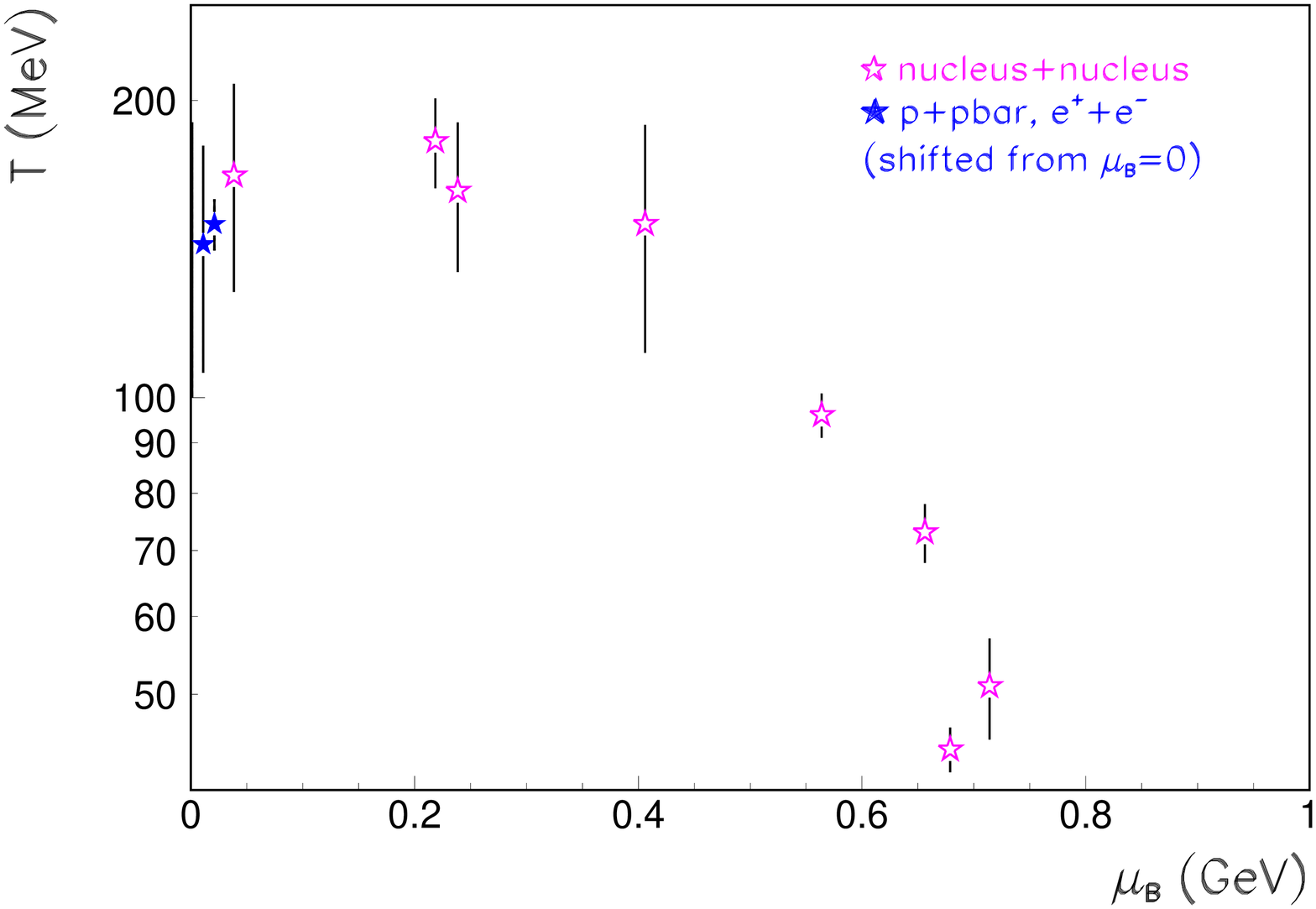,width=120mm}}
\end{center}
\caption{
The temperature as a function of
the baryochemical potential for several nucleus+nucleus, hadron+hadron and
lepton+lepton collisions.
We demand for the fits confidence level $>$ 10\%.
}
\label{t_vs_mub}
\vspace*{-0.5cm}
\end{figure}

\begin{figure}[ht]
\vspace*{-0.5cm}
\begin{center}
\mbox{\epsfig{file=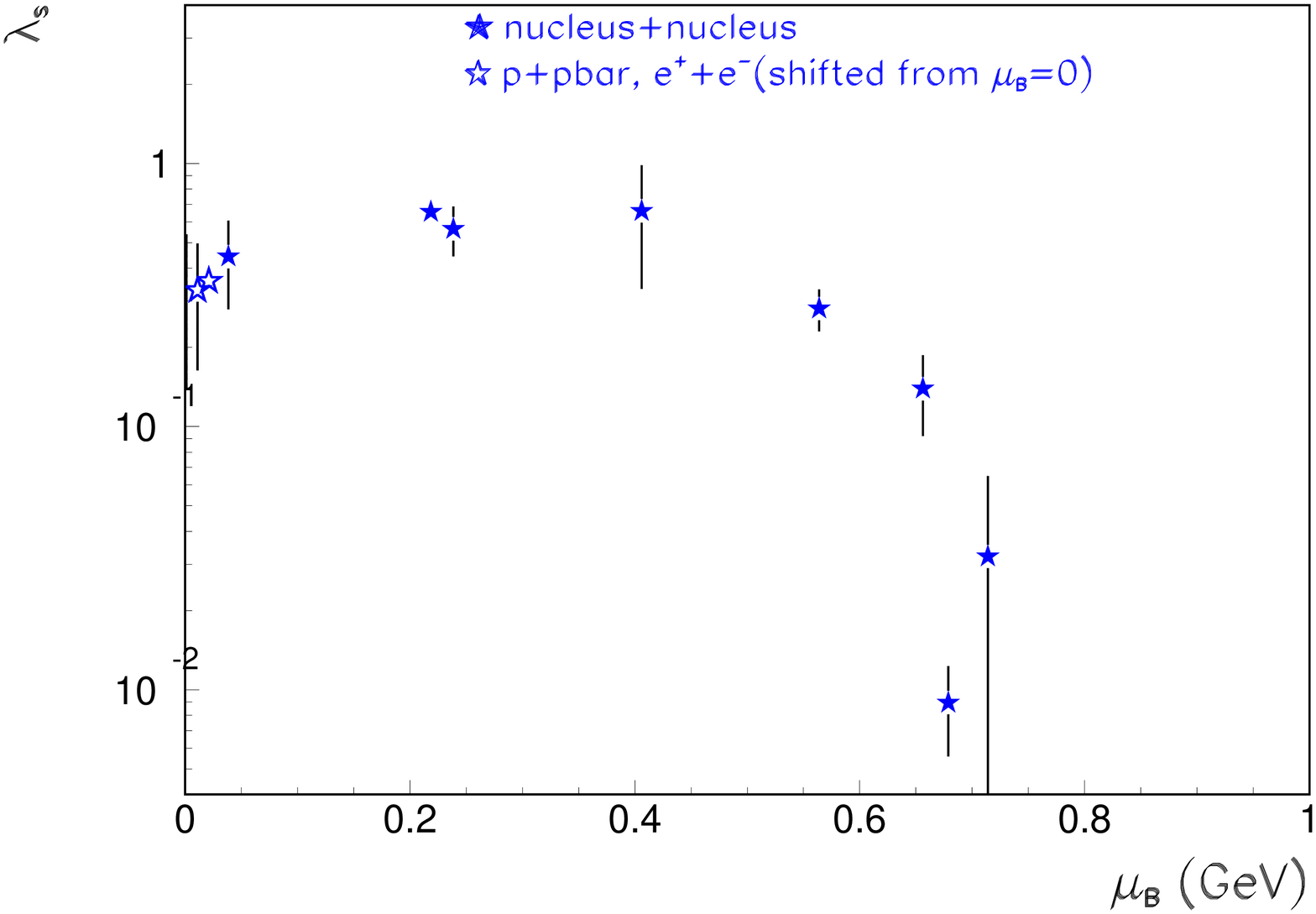,width=120mm}}
\end{center}
\caption{
The $\lambda_s$ factor as a function of
the baryochemical potential for several nucleus+nucleus, hadron+hadron and
lepton+lepton collisions.
We demand for the fits confidence level $>$ 10\%.
}
\label{ls_vs_mub}
\vspace*{-0.5cm}
\end{figure}

\begin{figure}[ht]
\vspace*{-0.5cm}
\begin{center}
\mbox{\epsfig{file=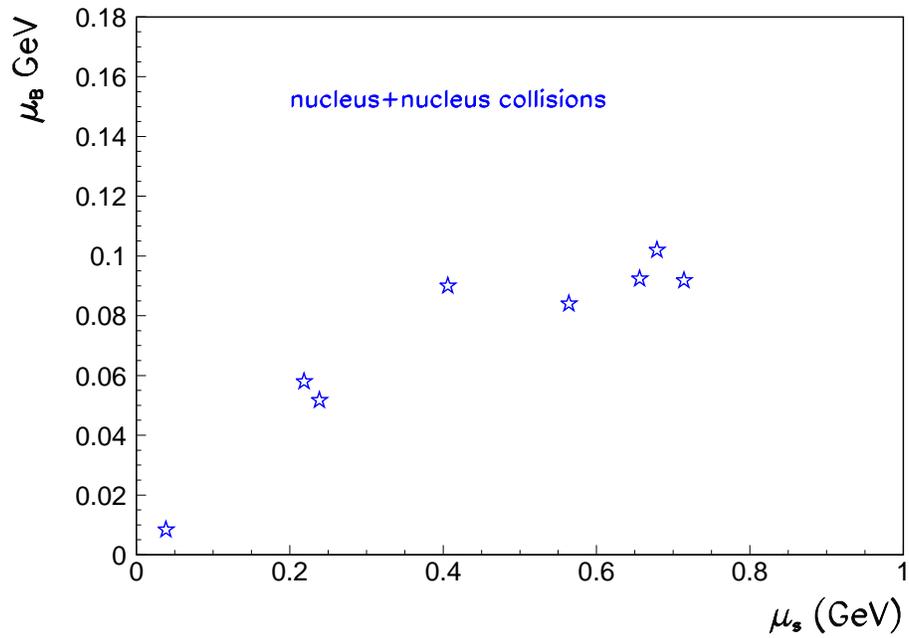,width=120mm}}
\end{center}
\caption{
The baryochemical potential as a function of
the strangeness chemical potential for several nucleus+nucleus
 collisions.
We demand for the fits confidence level $>$ 10\%.
}
\label{mub_vs_mus}
\vspace*{-0.5cm}
\end{figure}

\begin{figure}[ht]
\vspace*{-0.5cm}
\begin{center}
\mbox{\epsfig{file=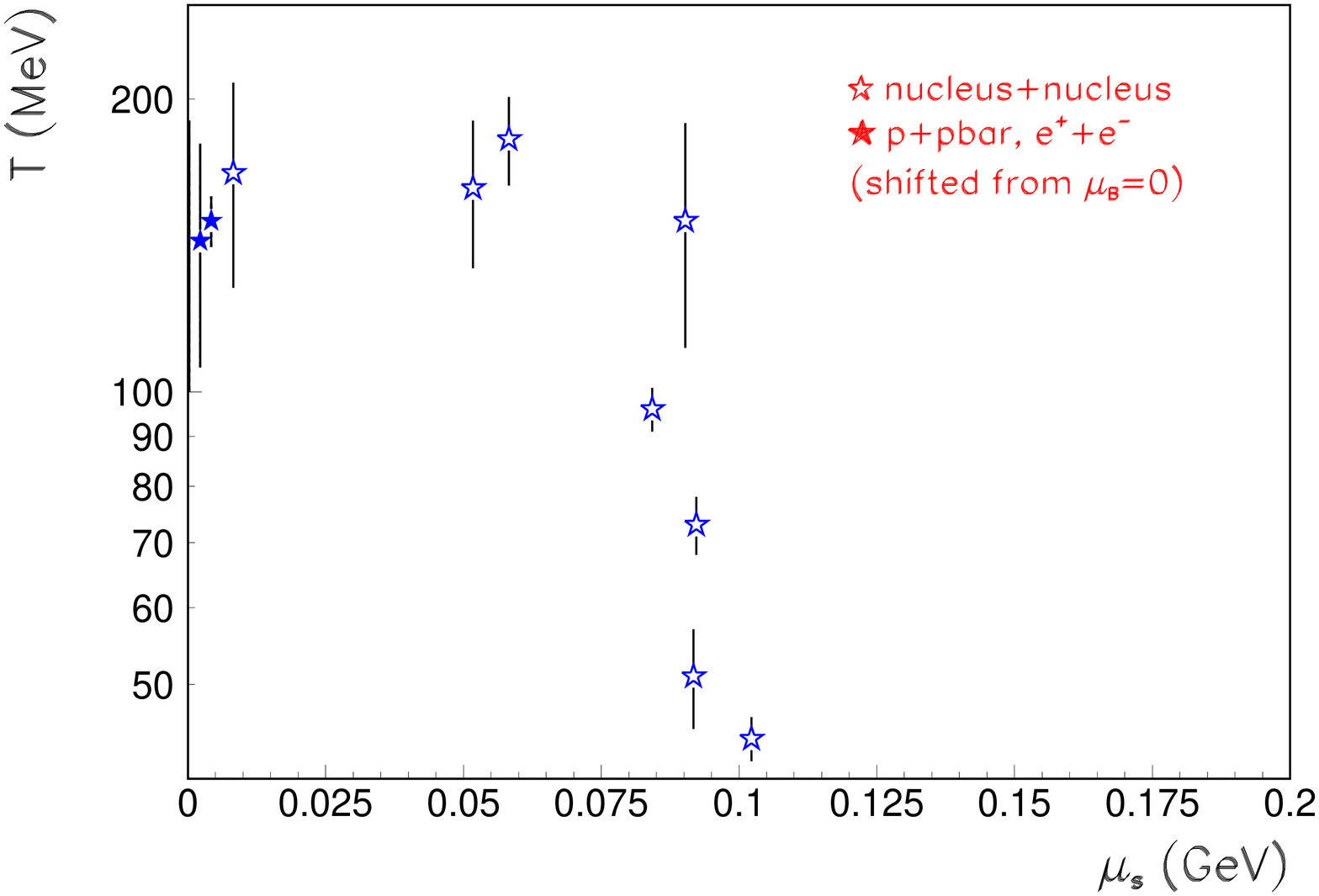,width=120mm}}
\end{center}
\caption{
The temperature as a function of
the strangeness chemical potential for several nucleus+nucleus
hadron+hadron and lepton+lepton collisions.
We demand for the fits confidence level $>$ 10\%.
}
\label{t_vs_mus}
\vspace*{-0.5cm}
\end{figure}

\begin{figure}[ht]
\vspace*{-0.5cm}
\begin{center}
\mbox{\epsfig{file=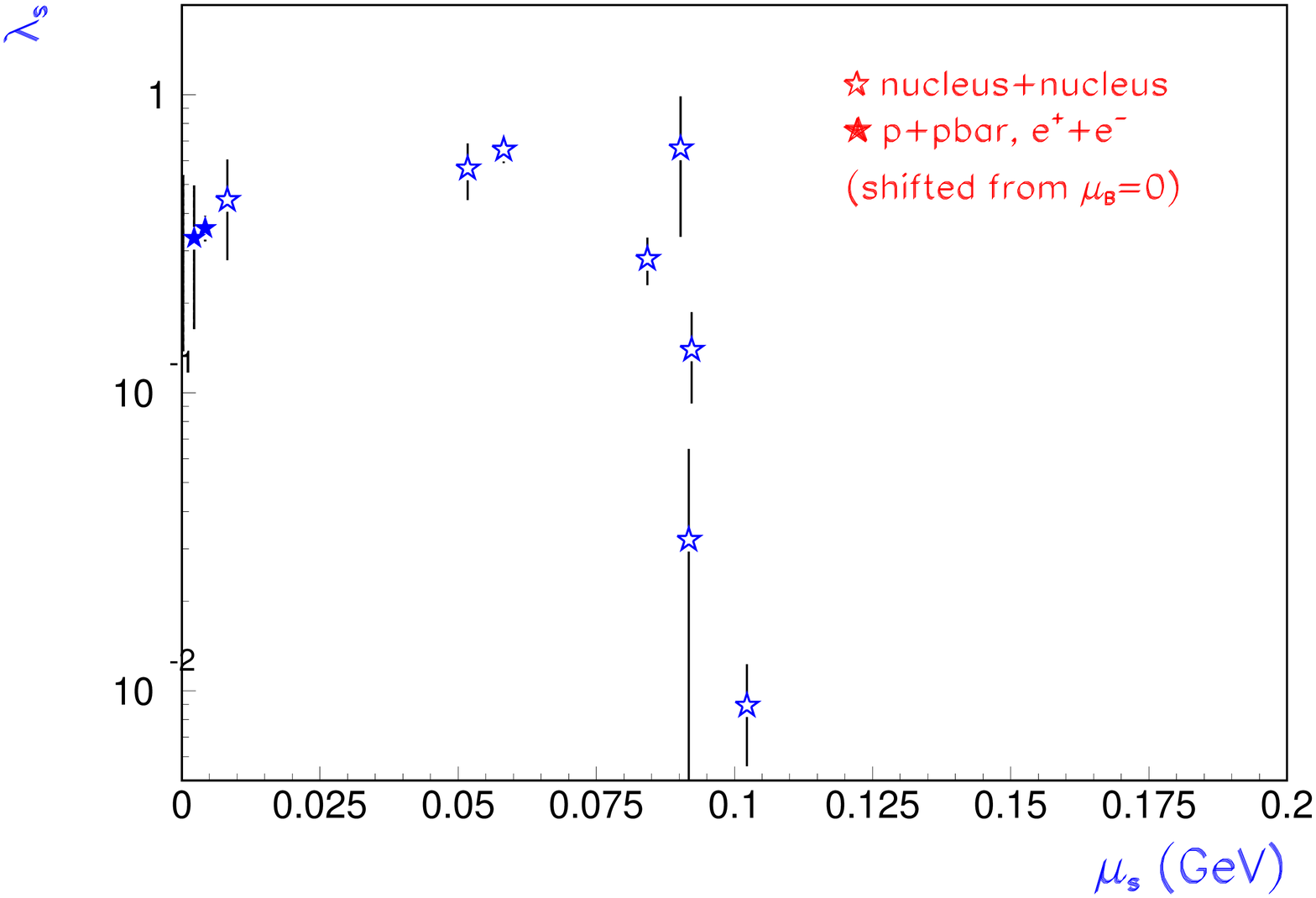,width=120mm}}
\end{center}
\caption{
The $\lambda_s$ factor as a function of
the strangeness chemical potential for several nucleus+nucleus
hadron+hadron and lepton+lepton collisions.
We demand for the fits confidence level $>$ 10\%.
}
\label{ls_vs_mus}
\vspace*{-0.5cm}
\end{figure}

\noindent
The correlation between the extracted thermodynamic parameters
temperature ($T$), strangeness suppression factor
($\lambda_s$)
and the chemical potentials of all systems together with results
from reference \cite{mapping}, 
is shown in the figures
\ref{t_vs_mub}, \ref{ls_vs_mub},
\ref{mub_vs_mus},
\ref{t_vs_mus},  \ref{ls_vs_mus}.
\\

\section{Initial energy density estimation}

\noindent
We estimate the initial energy density for the collision
systems studied in section 2 by taking
 the nuclear energy density of two overlapping nuclei
 2 $\epsilon_A$ times the $\gamma$ factor of the colliding particles
 in the center of mass minus one:
  
  \begin{equation}
  \epsilon_{\gamma} \ = \ 2 \ \epsilon_A \ (\gamma -1)
  \label{gamma}
  \end{equation}
   
\noindent
with $\gamma \ = \ (\sqrt{s}/2)/m_{nucleon}$, and
$\epsilon_{A}$=0.138 GeV/fm$^3$ 
is the normal nuclear matter density.
The value in equation \ref{gamma}  multiplied by the stopping power
gives an estimate of the initial energy density available for heating.
This formula is better suited for low energy data for example AGS
where the applicability of the Bjorken formula \cite{bjorkenform} for the 
energy density calculation is questionable.
For a discussion of the systematic error of $\sim$ 30-50\% on the initial
energy density calculation for nucleus nucleus and particle collisions
respectively,
see reference \cite{mapping,hepph0004138}.
\vspace{0.3cm}
The resulting initial energy densities are:
\vspace{0.3cm}

 $\epsilon_{\ in}$(Au+Au $\sqrt{s}$=2.3 GeV) =  0.12 GeV/$fm^3$,
 \vspace{0.3cm}

  $\epsilon_{\ in}$(Au+Au $\sqrt{s}$=3.3 GeV) =  0.21 GeV/$fm^3$,
  \vspace{0.3cm}

$\epsilon_{\ in}$(Pb+Pb $\sqrt{s}$=8.8 GeV) =  1.01 GeV/$fm^3$,
\vspace{0.3cm}

\section{The energy density dependence of $\lambda_s$ at finite $\mu$}

\noindent
Two general comments are important for the understanding of the
behaviour of strangeness in particle and nuclear collisions
and our later discussion: 
Since collisions of protons and collisions of nuclei
at the same energy per nucleon do reach different
initial energy densities, it is proper to compare 
hadronic observables as a function of the initial energy density
instead of  $\sqrt{s}$.
\\
It is also a fact that 
in all investigated colliding systems with a initial nonzero
net baryon number,
the baryon and strangeness chemical  potentials
are nonzero and different.
Next to the initial baryon number, also the 
energy of the collisions and hence the stopping 
influences the baryochemical potential of the
final hadrons produced.
Therefore the final state of the same nuclei, when colliding at different
energy, will be described by a different baryochemical potential.
\\
This leads to the conclusions that a) the behaviour of strangeness
as a QGP signature can be discussed while comparing
systems with the same baryon and strangeness chemical potential
and  b) the initial energy density is a better 'critical' parameter
than $\sqrt{s}$, against which different colliding systems can be compared.
We take these as starting
points and study their consequences in the following discussion.
\\

\noindent
When comparing the strange to non strange particle ratio in
nucleus nucleus collisions with p+p collisions at the same energy
an enhancement is seen, which
increases with decreasing energy \cite{enhancement}.
This observation is quantified in
figure \ref{marek} taken from reference \cite{na49_kpi}, which
shows the ratio of ($\Lambda + \overline{\Lambda} + K)/\pi$
as a function of the variable F which is a function of the
$\sqrt{s}$ of the collision.
The line shows a model prediction from reference \cite{marek}
assuming that the phase transition occurs at the vicinity 
of the maximum.
(For recent literature on the interpretation of data on
strangeness see e.g. \cite{inter}.)

\begin{figure}[ht]
\vspace*{-0.5cm}
\begin{center}
\mbox{\epsfig{file=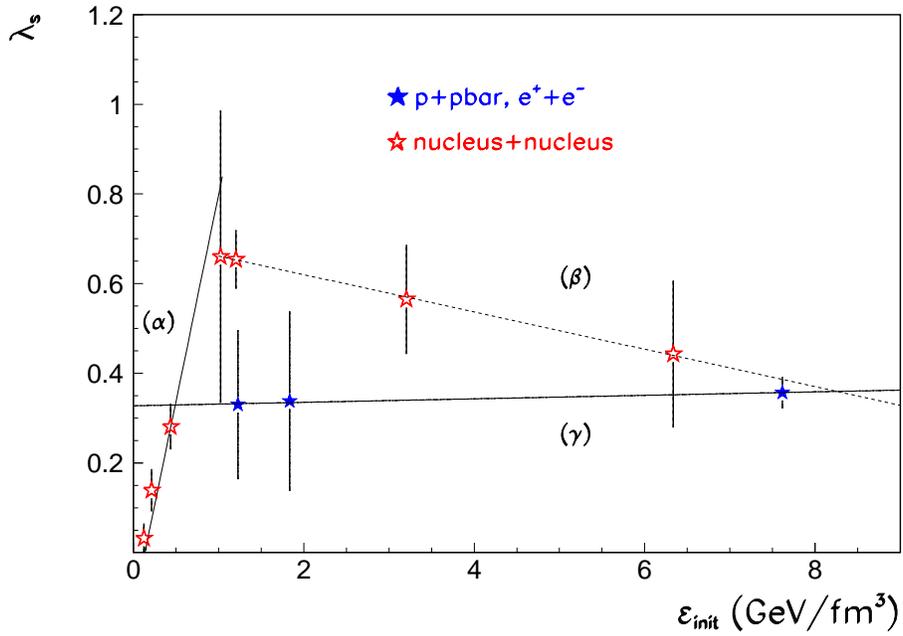,width=120mm}}
\end{center}
\caption{
The $\lambda_s$ factor as a function of
the initial energy density for several nucleus+nucleus, hadron+hadron and
lepton+lepton collisions.
We demand for the fits confidence level $>$ 10\%.
The lines $\alpha$ and $\beta$ show $\lambda_s$ at
nonzero $\mu_B$, while the line $\gamma$
show $\lambda_s$ at zero $\mu_B$.
}
\label{mub}
\vspace*{-0.5cm}
\end{figure}

\noindent
In figure \ref{mub} the strangeness suppression factor $\lambda_s$
is shown as a function of the initial energy density.
The open stars show nucleus nucleus collision results 
from the present analysis and the ones from reference 
\cite{mapping}, while the closed stars show results from
$p \overline{p}$ and $e^+ e^-$ collisions from
reference \cite{mapping}.
The lines shown are linear fits to the data points
to illustrate the tendency of the data.
\\
The  $\lambda_s$ points of all investigated nuclear collisions, 
define the  $\lambda_s$
behaviour at non zero and varying chemical potentials
(the two lines ($\alpha$) and ($\beta$)
of the  upper half of the triangle shown).
The horizontal line ($\gamma$) going through the $p \overline{p}$ and $e^+ e^-$  data
defines the  $\lambda_s$ value at the zero chemical potentials
in the $\epsilon_i$ region shown.

\noindent
The $\lambda_s$ factor grows reaching a maximum around
$\epsilon_i$ $\sim$ 1 GeV/fm$^3$ 
and then decreases continuously and almost linearly towards an asymptotic value
which is defined by the curve with zero chemical potentials
of the $p \overline{p}$ and $e^+ e^-$ data.
Therefore it follows	 that no saturation of the $\lambda_s$ value
is reached at energy density between 1 GeV/$fm^3$ and 
8 GeV/$fm^3$ in nucleus nucleus collisions, 
as the one indicated in figure \ref{marek}.
\\

\noindent
We will argue that the phenomenon of enhancement of the
$\lambda_s$ value in nucleus collisions
as compared to the $\lambda_s$ values in elementary particle
collisions,  quantified in figure \ref{mub}
by the area of the triangle,
is mainly due to the difference in the baryochemical potential.
\\
With increasing energy, the central rapidity region in nuclear
collisions becomes increasingly net baryon free
approaching a state of a nucleus+antinucleus collision.
\\
This tendency is shown in figure \ref{mub}
by the decrease of the line ($\alpha$) with increasing $\epsilon_i$.
The statistical significance of the present data does not allow
a thorough investigation of the exact shape of the decrease.
In particular in order to find deviations from a linear behaviour,
more data are needed.
\\
The nonzero potential $\lambda_s$ line ($\alpha$) crosses the
zero potential $\lambda_s$ horizontal line ($\gamma$) at an
initial energy density of $\sim$ 8-9 GeV/$fm^3$ (depending
on the way we extrapolate e.g. using a linear, exponential or a polynomial
distribution).
The linear fit crosses at 8.25 GeV/$fm^3$.
\\
This indicates that the limiting value of $\lambda_s$ (and an
almost net baryonfree midrapidity region) will
be achieved at nuclear collisions reaching an initial energy density
of $\sim$ 8-9 GeV/$fm^3$.\footnote{The  net baryonfree midrapidity region in a
particle+particle collision and its equivalence with
particle+antiparticle collision, can probably be
achieved only in the limit of infinite energy.
We address here the question at which energy 
this asymptotic value is significantly approached 
within the errors.}
Assuming that $\epsilon_i$ scales as the logarithm of 
$\sqrt{s}$, we find that the $\sqrt{s}$ needed to achieve
the limiting value of $\lambda_s$ using nucleus nucleus
collisions is approximately 3-8 TeV.
This energy density could therefore be achieved by the Large Hadron Collider
at CERN.

\section{The energy density dependence of $\lambda_s$ at zero $\mu$}

\begin{figure}[ht]
\begin{center}
\mbox{\epsfig{file=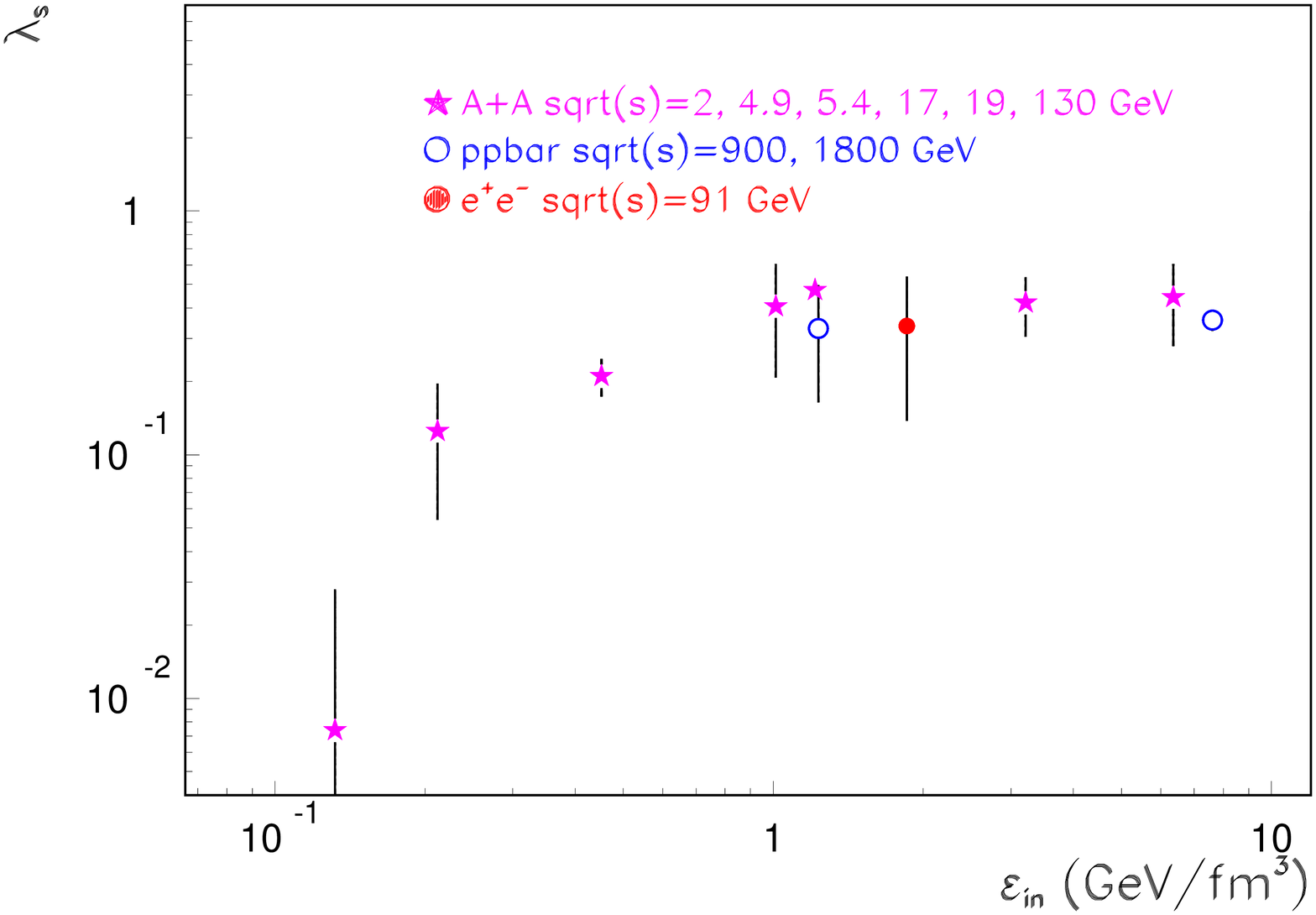,width=120mm}}
\end{center}
\caption{
The $\lambda_s$ factor (in logarithmic scale)
extrapolated to zero fugacities along
an isentropic path,
as a function of
the initial energy density for several nucleus+nucleus, hadron+hadron and
lepton+lepton collisions.
We demand for the fits confidence level $>$ 10\%.
}
\label{0}
\end{figure}

\begin{figure}[ht]
\begin{center}
\mbox{\epsfig{file=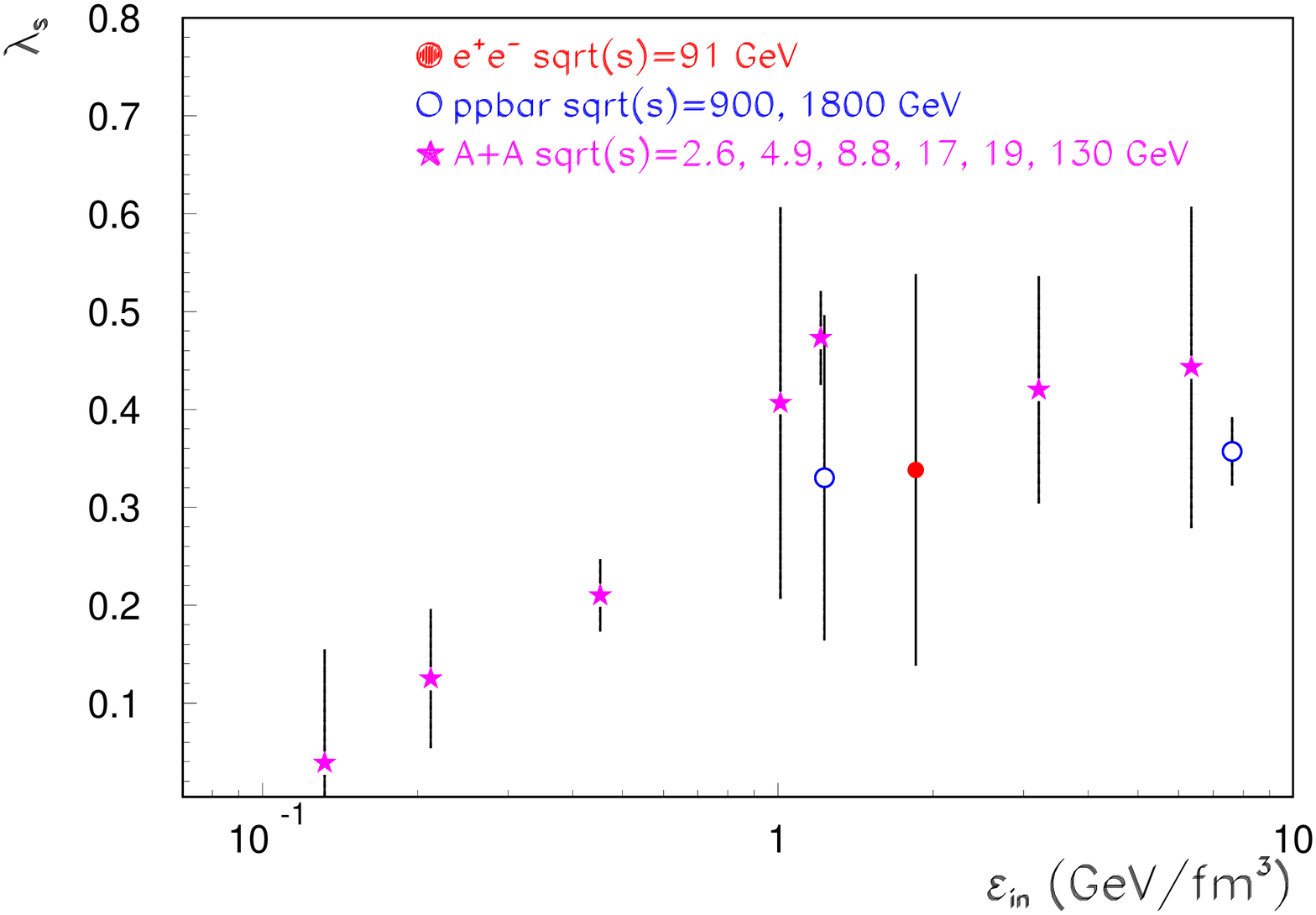,width=120mm}}
\end{center}
\caption{
The $\lambda_s$ factor (in linear scale)
extrapolated to zero fugacities along
an isentropic path,
as a function of
the initial energy density for several nucleus+nucleus, hadron+hadron and
lepton+lepton collisions.
We demand for the fits confidence level $>$ 10\%.
}
\label{0_lin}
\end{figure}

\noindent
In the following we discuss the 
$\epsilon_i$ dependence of the $\lambda_s$ factor
after extrapolating all thermodynamic
states with non zero baryochemical potential to zero,
along an isentropic path.
The result is shown in figure \ref{0} in logarithmic representation
(to show the point of smallest $\lambda_s$),
and in figure \ref{0_lin} in linear representation.
All $\lambda_s$ values show a universal behaviour
increasing from below until $\epsilon_i$ $\sim$ 1 GeV/fm$^3$
and then saturating.
\\

\noindent
Several comments can be made:
\\
\noindent
1) The figure  clearly demonstrates that the peak
of $\lambda_s$ and of the $K/\pi$ ratio
at 40 A GeV Pb+Pb collisions and their drop towards higher energies,
seen in figures \ref{marek} and \ref{mub} 
is due to the nonzero baryochemical potential in the  collisions.
The $\lambda_s$ factor is the same for Pb+Pb
collisions at 40 and at 158 A GeV in figure \ref{0}
(4th and 8th points from the left).
\\
Therefore, the 
so called 'strangeness suppression' phenomenon \cite{na49_kpi,redlichqm2001},
that is, the decrease
of the strange to non strange particle ratios (e.g. $K/ \pi$)
from 40 GeV Pb+Pb towards 158 GeV Pb+Pb collisions
is explained by the
different baryochemical potentials of these systems.
\\
This interpretation is supported by the difference seen in the $\sqrt{s}$
dependence of the $K^+/\pi^+$ and the $K^-/\pi^-$ ratio
as a function of $\sqrt{s}$ \cite{na49_kpi}.
In particular it is seen that the peak at 40 A GeV Pb+Pb
colisions, appears only in the
$K^+/\pi^+$ ratio ($K^+$ is 'forced' by the high $\mu_B$
to be abudantly produced in association with $\Lambda$).
\\

\noindent
2) The increase of the double ratio $K/\pi$ (A+A/p+p)
with decreasing $\sqrt{s}$ discussed e.g. in the review talk
\cite{ogilvie_qm2001} 
is a natural consequence arising from the above ideas.\footnote{The
increase is of course large when the
threshold for strangeness production in p+p reactions is approached,
while subthreshold strangeness 
production can still occur in nuclear collisions, but
this limiting case is not the main point of the present discussion.}
\\
From this comparison the apparent strangeness enhancement
is increasing towards   lower energies.
However the comparison leading to this conclusion is not 
taking into account the varying characteristic parameters.
\\

\noindent
3)
The $K/\pi$ ratio
is investigated in reference \cite{redlichqm2001}
and shows the above mentioned strangeness suppression
with $\sqrt{s}$,
when constructed using yields of particles in the full  phase space acceptance,
while it remains constant at midrapidity.
\\
This can be explained by the fact that
the baryochemical potential reaches its lowest value
at the midrapidity of nucleus nucleus collisions
and the highest one in the forward and backward rapidity regions
(multiple reservoirs) e.g. \cite{Sonja2,myphd}.
Therefore, when considering  $K/\pi$ at midrapidity
the bias from nonzero $\mu_B$ 
is minimized as compared to full acceptance yields.
\\

\noindent
4) The enhancement seen in strange to non strange particle
yields
in central p+A collisions when compared to p+p at the same energy
\cite{pA},
is  explained in the same way, since central p+A collisions
reach a higher $\mu$ and a 
higher initial energy density as p+p collisions at the
same $\sqrt{s}$.
\\

\noindent
5) The  strangeness enhancement
as defined in the literature (double ratio of 
$\lambda_s$ in  A+A over p+p collisions at the same $\sqrt{s}$)
and as illustrated in figure \ref{marek} 
finds the same explanation.
\\
Normalizing to the same $\mu_B$ value (here choosen to be 
zero for simplicity) and  comparing at the same
initial energy density,
the above mentioned 'strangeness enhancement' dissapears.
\\
The $\lambda_s$ factor for particle collisions
is within the error compatible with nucleus nucleus collisions
as seen in figures \ref{0} and \ref{0_lin}.
\\

\begin{figure}[ht]
\begin{center}
\mbox{\epsfig{file=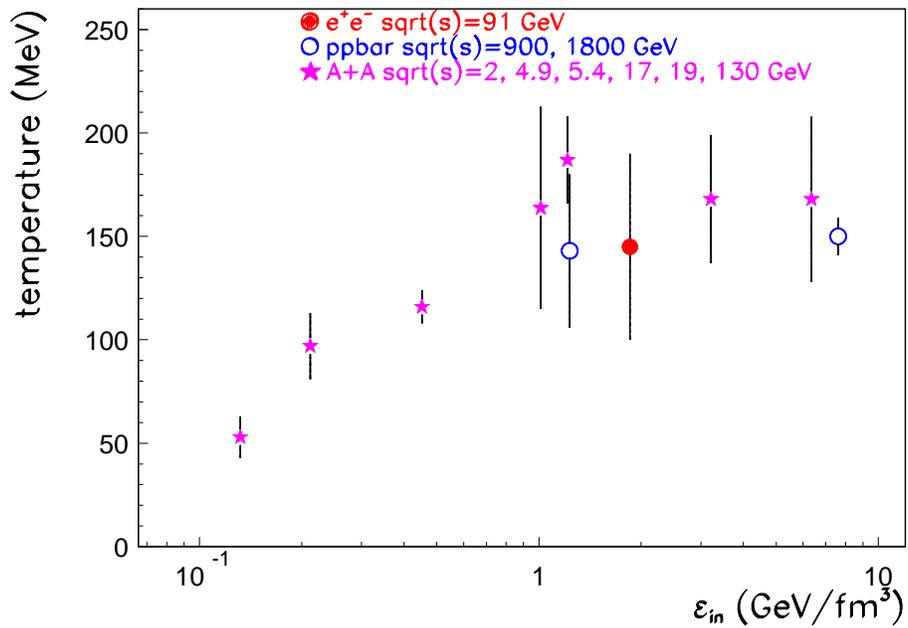,width=120mm}}
\end{center}
\caption{
The temperature extrapolated to zero fugacities along
an isentropic path,
as a function of
the initial energy density for several nucleus+nucleus, hadron+hadron and
lepton+lepton collisions.
We demand for the fits confidence level $>$ 10\%.
}
\label{t}
\end{figure}

\noindent
6) 
Enfin: what can we learn from strangeness production
about the QCD phase transition ?
\\
First, figure \ref{0} exhibits a very distinct feature of
$\lambda_s$ as a function of $\epsilon_i$:
$\lambda_s$ is increasing 
with 
initial energy density increasing  from 0.14 to 1 GeV/$fm^3$.
After this $\epsilon_i$ value, $\lambda_s$ saturates at a limiting
value of $\lambda_{\ lim} \ =  0.365 \pm 0.033 \pm 0.07$ \cite{mapping}.
\\
Second, this behaviour is followed universally by all
collisions studied.
\\
\noindent
Third,   $\lambda_s$ is following
closely the $\epsilon_i$  dependence of the temperature
at the chemical freeze out of hadrons
(as  shown in figure \ref{t}).
\\

\noindent
Assuming that local equilibrium conditions are
reached in all collisions studied, 
and therefore that temperature is defined,
we  interpret the dramatic change of 
the temperature and of $\lambda_s$ at chemical freeze-out
near $\epsilon_i$ $\sim$ 1 GeV/fm$^3$,
 as critical behaviour:
 the onset of saturation occurs when
the critical energy density for the QCD phase transition
is reached.
\\
\noindent
The $T$  as well as
$\lambda_s$ would continue to rise  with $\epsilon_i$ if no 
phase transition occurs.
\\

\noindent
A bias in this interpretation is the  implicit assumption
that  $T$ does not cool down
without noticing the transition
due to some peculiar expansion dynamics and its $\sqrt{s}$
dependence.
However in this  case the cooling mechanism should 
generate the  universal behaviour
seen in figures \ref{0} and \ref{t} by coincidence.

\noindent
Furthermore, since pressure and expansion 
characteristics  have been measured to be different
in different A+A collisions (e.g. there are strong
flow phenomena pointing to high initial pressure in A+A
collisions which are more pronounced at RHIC than
in SPS, and not seen in p+p collisions)
it seems difficult to 
explain a universally flat temperature curve exhibited by
all p+p and A+A data,
by a non universal expansion dynamics.
\\

\noindent
A phenomenon analogous to the one seen in figure \ref{t}
is the following:
\\
We fill a box with water and look for the
 water-vapour phase transition
 without tools to detect vapour.
 Each time the transition to vapour (=QGP)
occurs
we thus wait until the vapour condensates back to water (=hadron gas), 
in order  to measure its temperature.
\\
We make a plot of the water  temperature
as a function of the applied heat, and it looks as figure
\ref{t}, namely it rises and saturates
at the value of $\leq$ 100$^o$ Celsius.
\\
Adding  salt to water and repeating the experiment would 
result in different critical values rising with salinity.
\\
The baryochemical potential is like salt for hadronic systems.
To achieve measuring one single curve one has to use
the same salinity, as we do in figures \ref{t} and \ref{0}.
As a result, the border of the QCD phase transition can be 
drawn 
and the critical energy density can be extracted 
selfconsistently from the data,
independent of any model  predicting where the boundary
must be.
\\

\noindent
The dependence of the temperature on the baryon and strangeness
chemical potentials
is shown in figures \ref{t_vs_mub} and \ref{t_vs_mus}.
\\
If all these points would lie exactly on the critical border,
their  extrapolation to equivalent states at zero potentials
would lead to one single critical temperature.
\\
If all points would not lie on the critical border
but well below, their extrapolation to the equivalent states
with $\mu_B$ zero would give scattered temperatures
below the critical one,  without showing signs of saturation.
\\

\noindent
One could counterargue 
that the saturation is caused by the freeze out conditions
and the critical temperature is never reached in the initial
state, (for example if there is no $T_{crit}$ nearby).
\\
In this case however the freeze out conditions should change
with the collision energy, because the cross sections
of particle interactions and the mean free path changes too
with $\sqrt{s}$.
Again saturation is then coincidental.
\\
A second argument against such coincidence is the following:
If the freeze out 
conditions do imply a universal temperature at chemical freeze out
then there should be no variation  in the chemical freeze out
temperature at zero chemical potentials.
That is the plot of figure \ref{t} should be completely flat
for all thermal systems.
\\
A third argument is
the similarity of the critical parameters
that we extract in the above described picture and in \cite{mapping}
with the expected critical parameters from QCD
($T_{crit} \sim 200 MeV$, $\epsilon_{crit} \sim 1 GeV/fm^3$
\cite{mapping} and from other calculations too
e.g. from lattice simulations \cite{lattice}).
Even if the estimation of the initial energy
density \cite{bjorkenform}  has a large inherent error, 
e.g. in the assumption of a formation time of 1 fm/c, still
the error hardly amounts to a factor 7 (the difference
between $\epsilon_i$ of Tevatron and the $\epsilon_i$=1 GeV/$fm^3$).
Also our limiting temperature determination is more reliable
than the initial energy
density estimate and yet $T_{lim}$ and $\epsilon_{crit}$
agree reasonably with each other
and with the QCD predictions.
Finally we conclude that Bjorken's estimate \cite{bjorkenform}
turns out to be correct.
\\

\noindent
7)
It is important to note, that while studing the
$\lambda_s$ or the temperature dependence on the
initial energy density using their values at non zero
chemical potential (figure \ref{mub})
one can not extract the critical 
energy density neither the  limiting temperature from
this study.
\\
In particular the values one can extract in this case,
simply reflect the fact
that we deal with a critical border which is a 2-dimensional
surface.
However more data are needed in order to observe this shift.
\\

\noindent
8)
All temperatures achieved with hadrons, are below the critical
temperature for sure, however they can approach the border.
We can measure only the approach to the critical surface from
the hadronic side, study 
the way the critical values are approached and
extract the critical parameters from critical behaviour.
\\

\begin{figure}[ht]
\begin{center}
\mbox{\epsfig{file=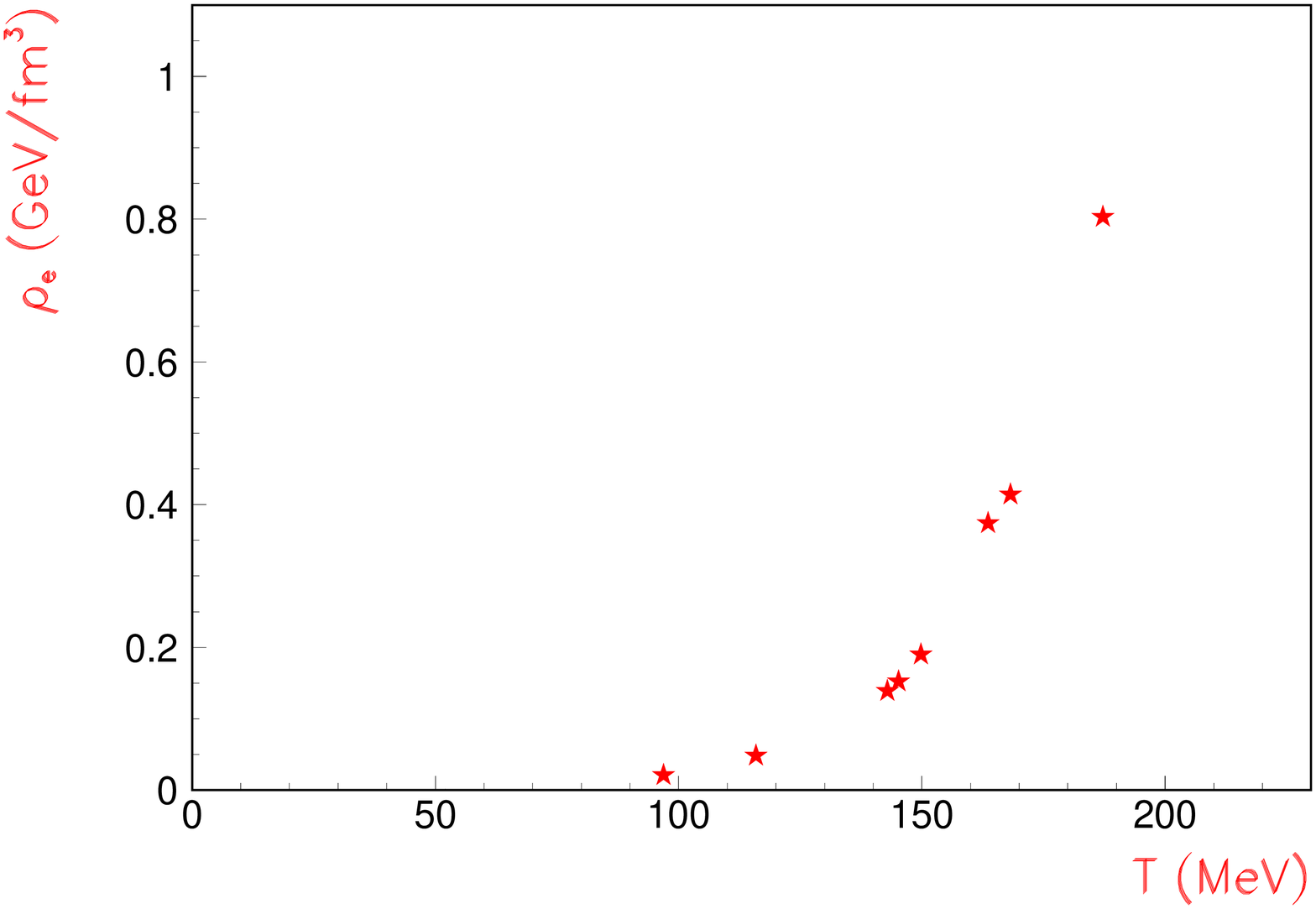,width=120mm}}
\end{center}
\caption{
The energy density $\rho_e$  as a function of the temperature
extrapolated to zero fugacities along an isentropic path
for many nucleus+nucleus, hadron+hadron and
lepton+lepton collisions.
We demand for the thermal model fits confidence level $>$ 10\%.
}
\label{rhoe_vs_t}
\end{figure}

\begin{figure}[ht]
\begin{center}
\mbox{\epsfig{file=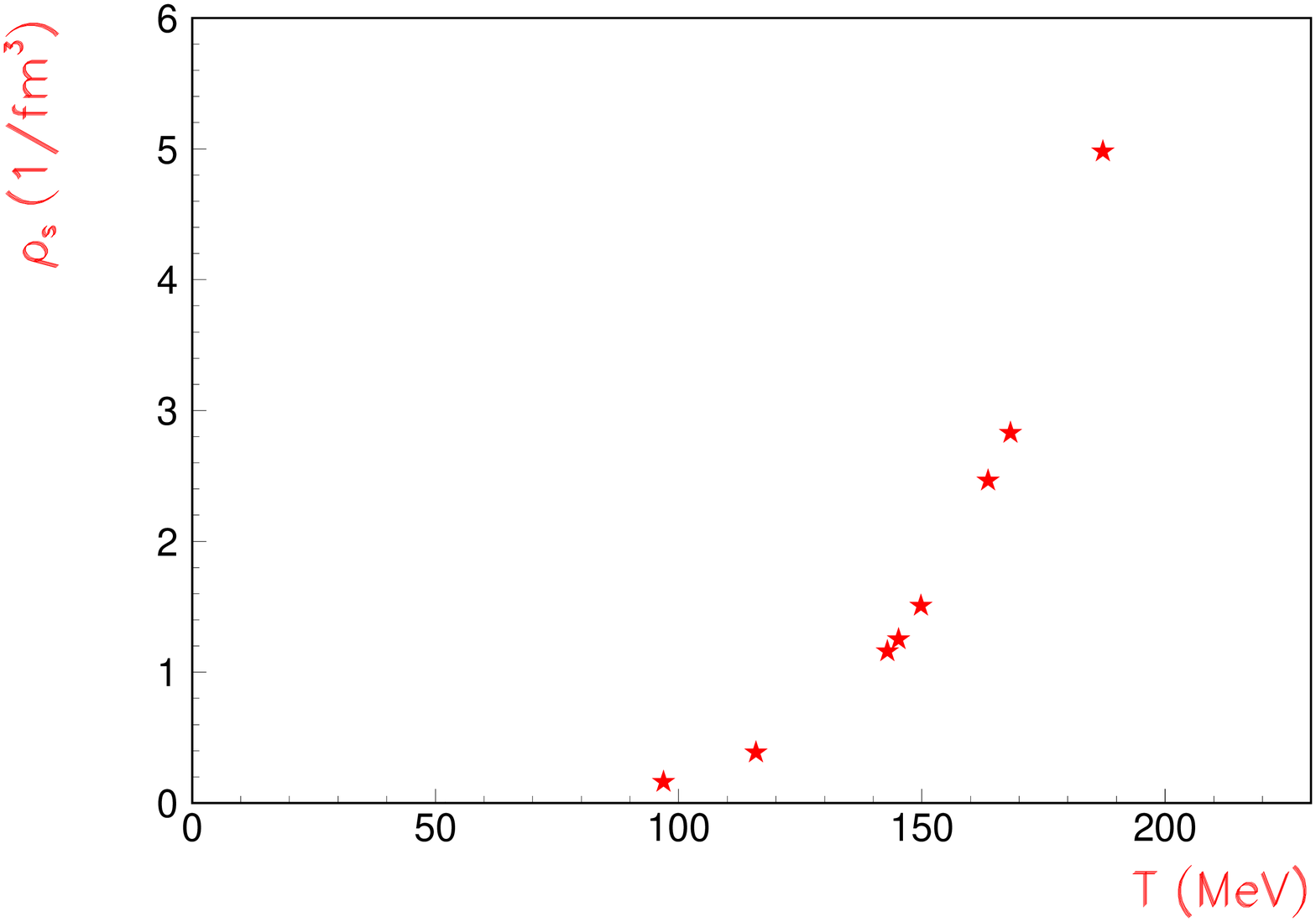,width=120mm}}
\end{center}
\caption{
The entropy density $\rho_s$  as a function of the temperature
extrapolated to zero fugacities along an isentropic path
for many nucleus+nucleus, hadron+hadron and
lepton+lepton collisions.
We demand for the thermal model fits confidence level $>$ 10\%.
}
\label{rhos_vs_t}
\end{figure}

\begin{figure}[ht]
\begin{center}
\mbox{\epsfig{file=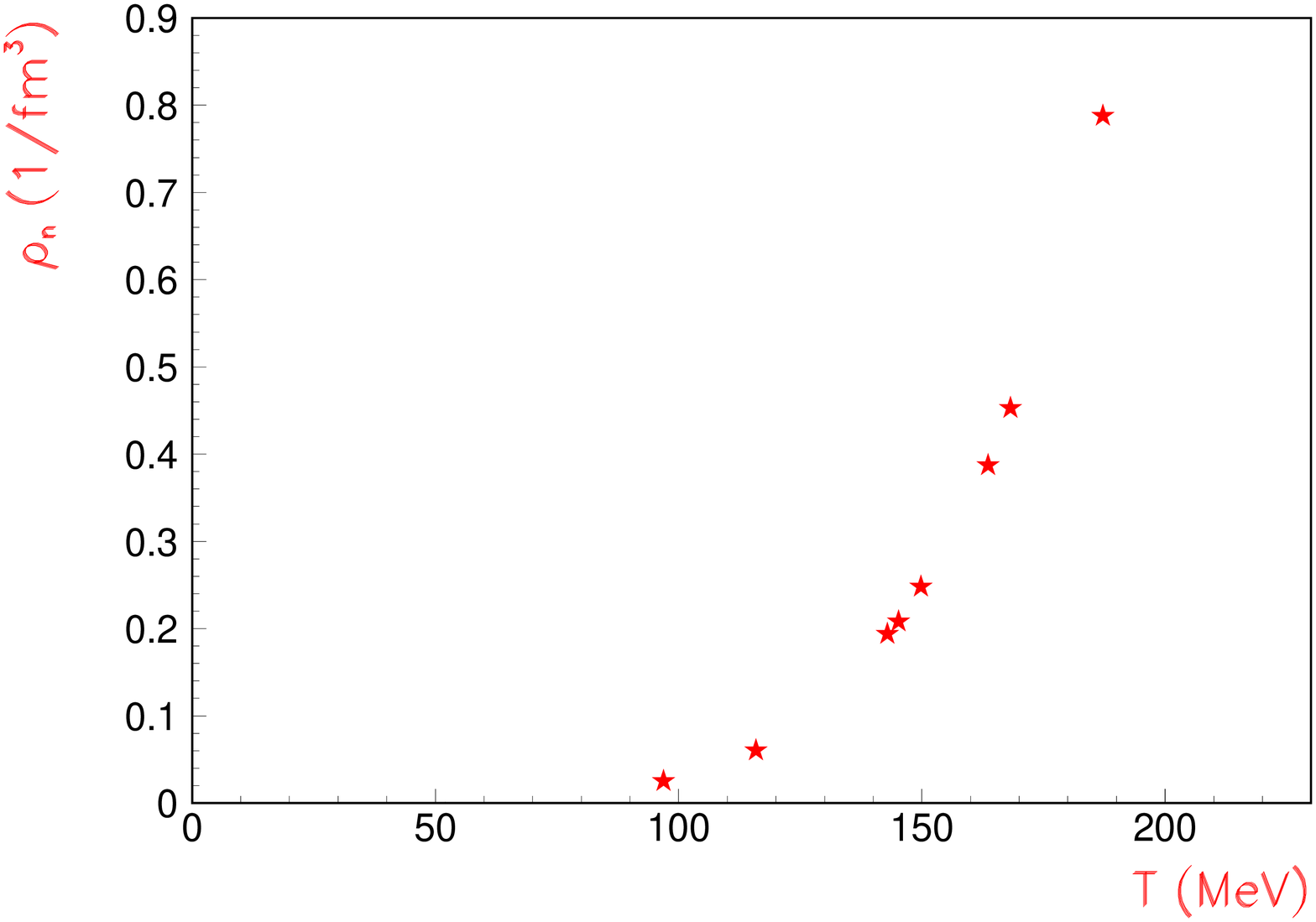,width=120mm}}
\end{center}
\caption{
The density $\rho_n$  as a function of the temperature
extrapolated to zero fugacities along an isentropic path
for many nucleus+nucleus, hadron+hadron and
lepton+lepton collisions.
We demand for the thermal model fits confidence level $>$ 10\%.
}
\label{rhon_vs_t}
\end{figure}

\begin{figure}[ht]
\begin{center}
\mbox{\epsfig{file=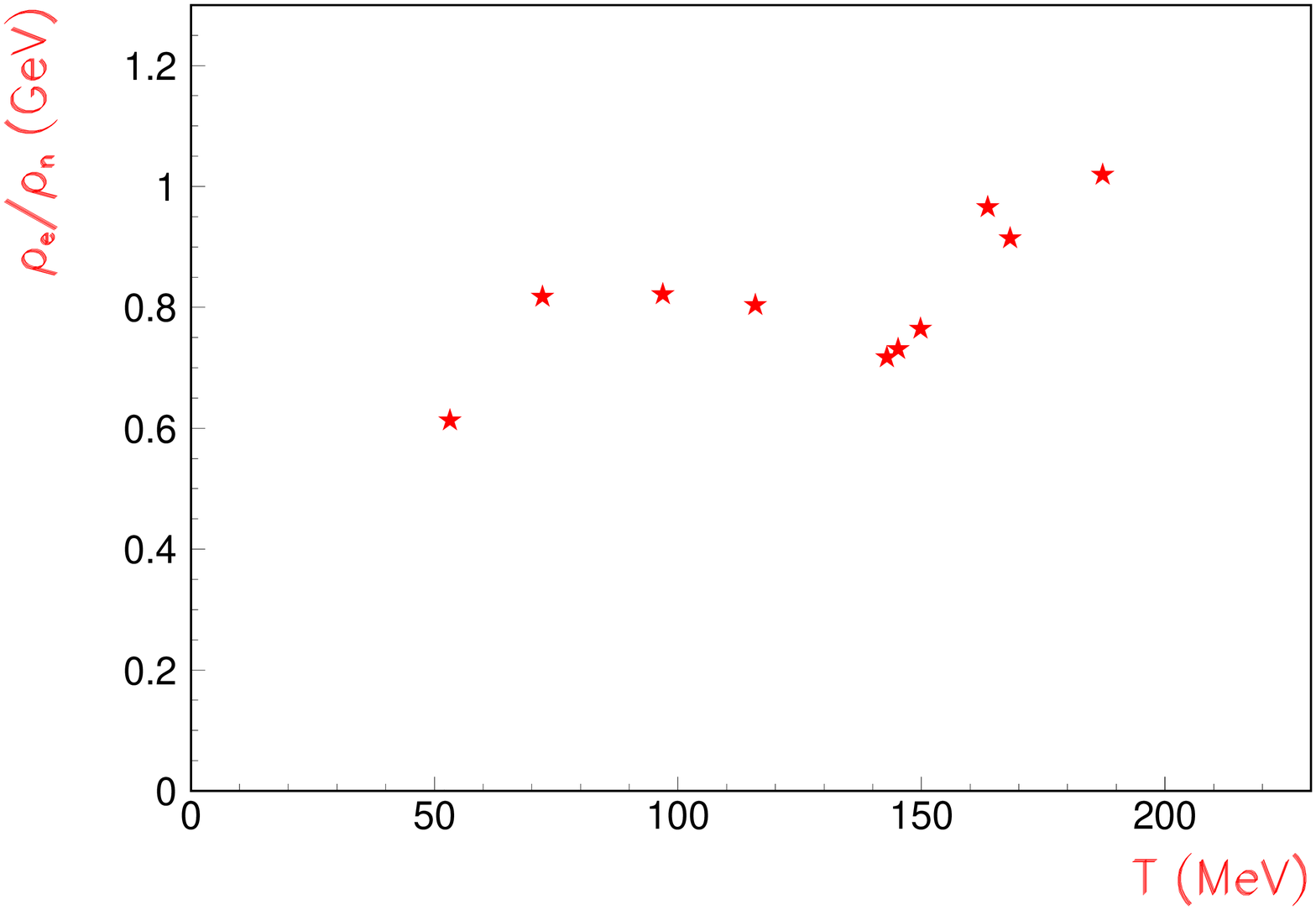,width=120mm}}
\end{center}
\caption{
The ratio of the energy density to the density 
as a function of the temperature
extrapolated to zero fugacities along an isentropic path
for many nucleus+nucleus, hadron+hadron and
lepton+lepton collisions.
We demand for the thermal model fits confidence level $>$ 10\%.
}
\label{rhoe_to_rhon_vs_t}
\end{figure}

\begin{figure}[ht]
\begin{center}
\mbox{\epsfig{file=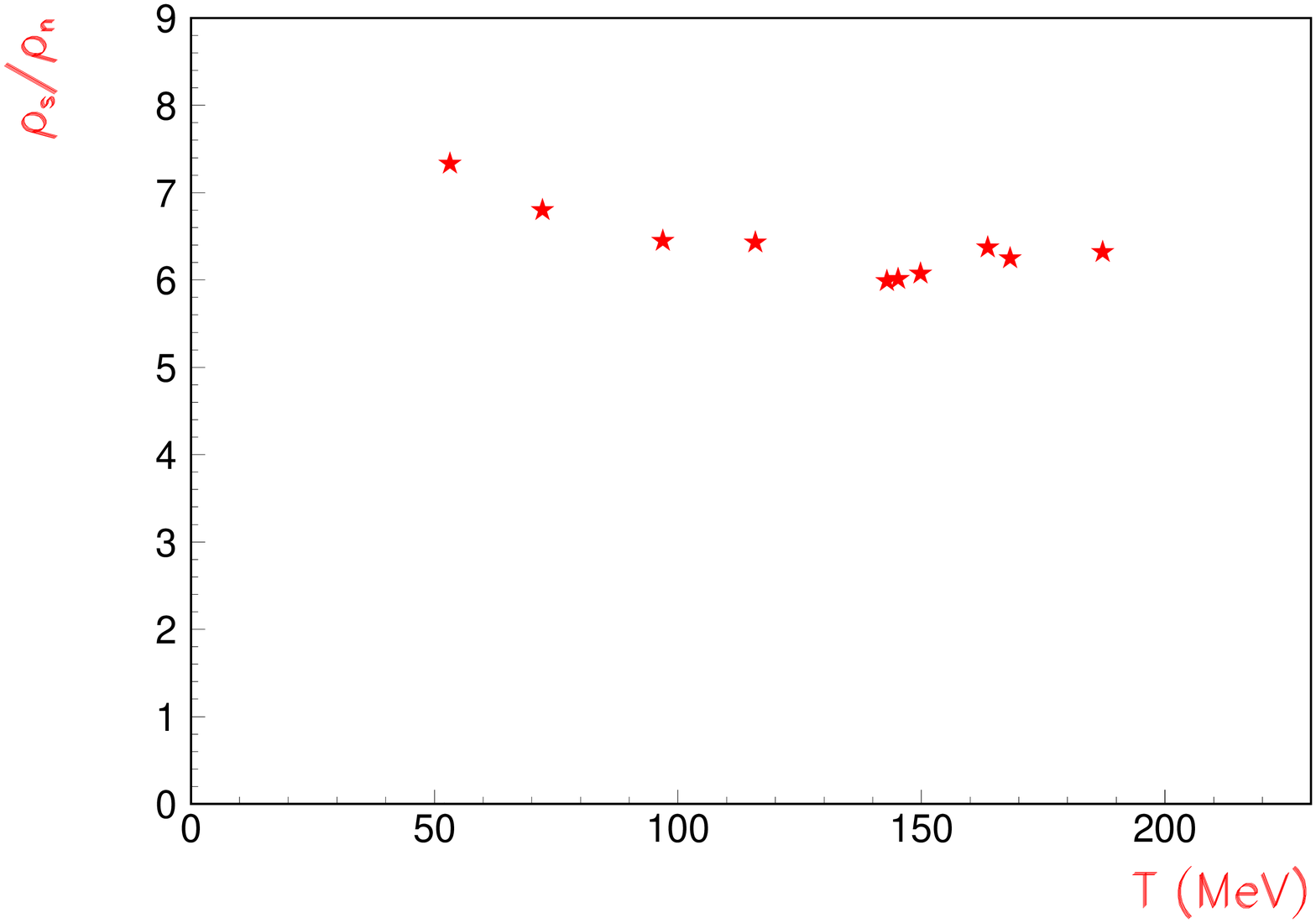,width=120mm}}
\end{center}
\caption{
The ratio of the entropy density to the density 
as a function of the temperature
extrapolated to zero fugacities along an isentropic path
for many nucleus+nucleus, hadron+hadron and
lepton+lepton collisions.
We demand for the thermal model fits confidence level $>$ 10\%.
}
\label{rhos_to_rhon_vs_t}
\end{figure}

\begin{figure}[ht]
\begin{center}
\mbox{\epsfig{file=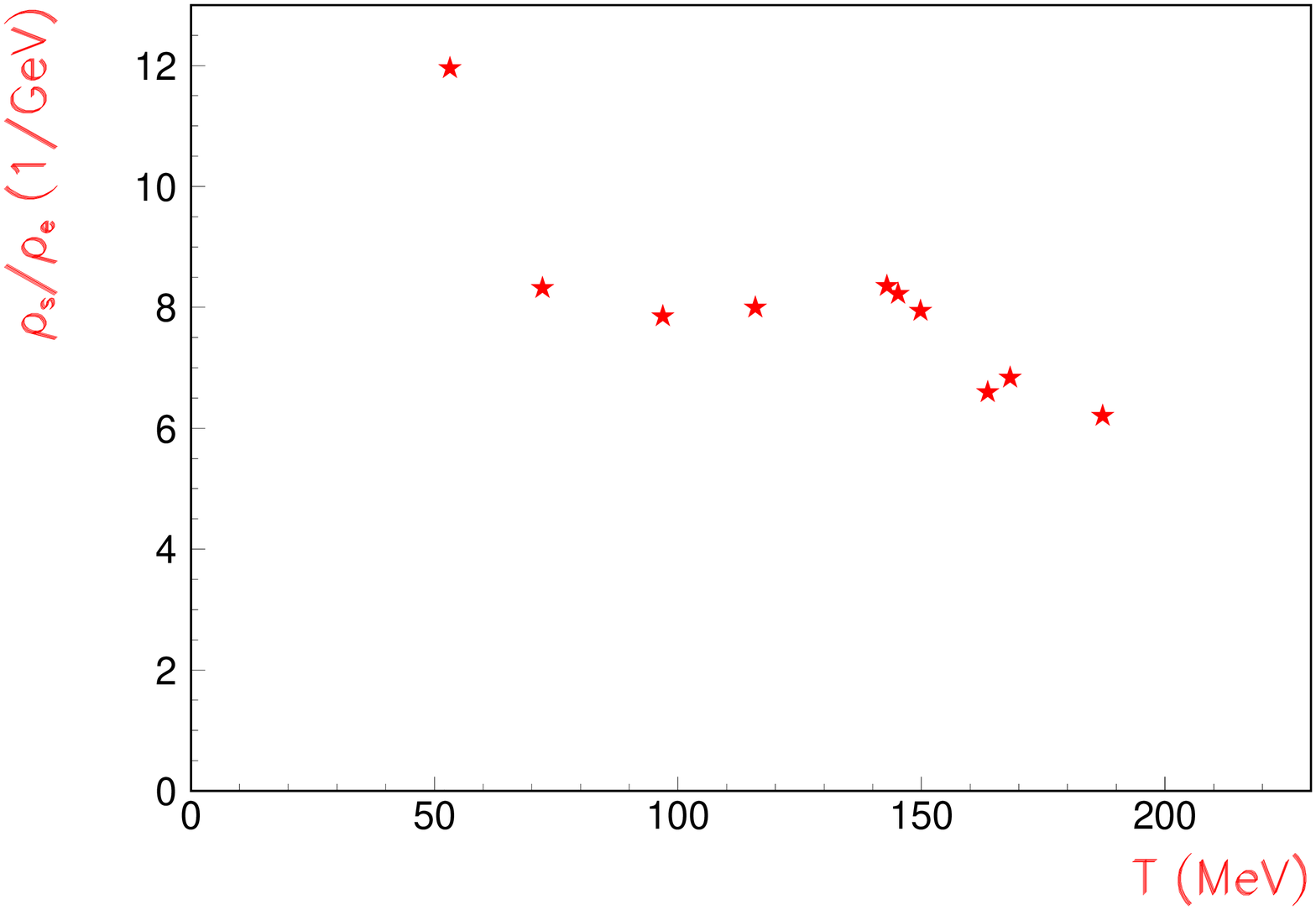,width=120mm}}
\end{center}
\caption{
The ratio of the entropy density to the energy density 
as a function of the temperature
extrapolated to zero fugacities along an isentropic path
for many nucleus+nucleus, hadron+hadron and
lepton+lepton collisions.
We demand for the thermal model fits confidence level $>$ 10\%.
}
\label{rhos_to_rhoe_vs_t}
\end{figure}

\begin{figure}[ht]
\begin{center}
\mbox{\epsfig{file=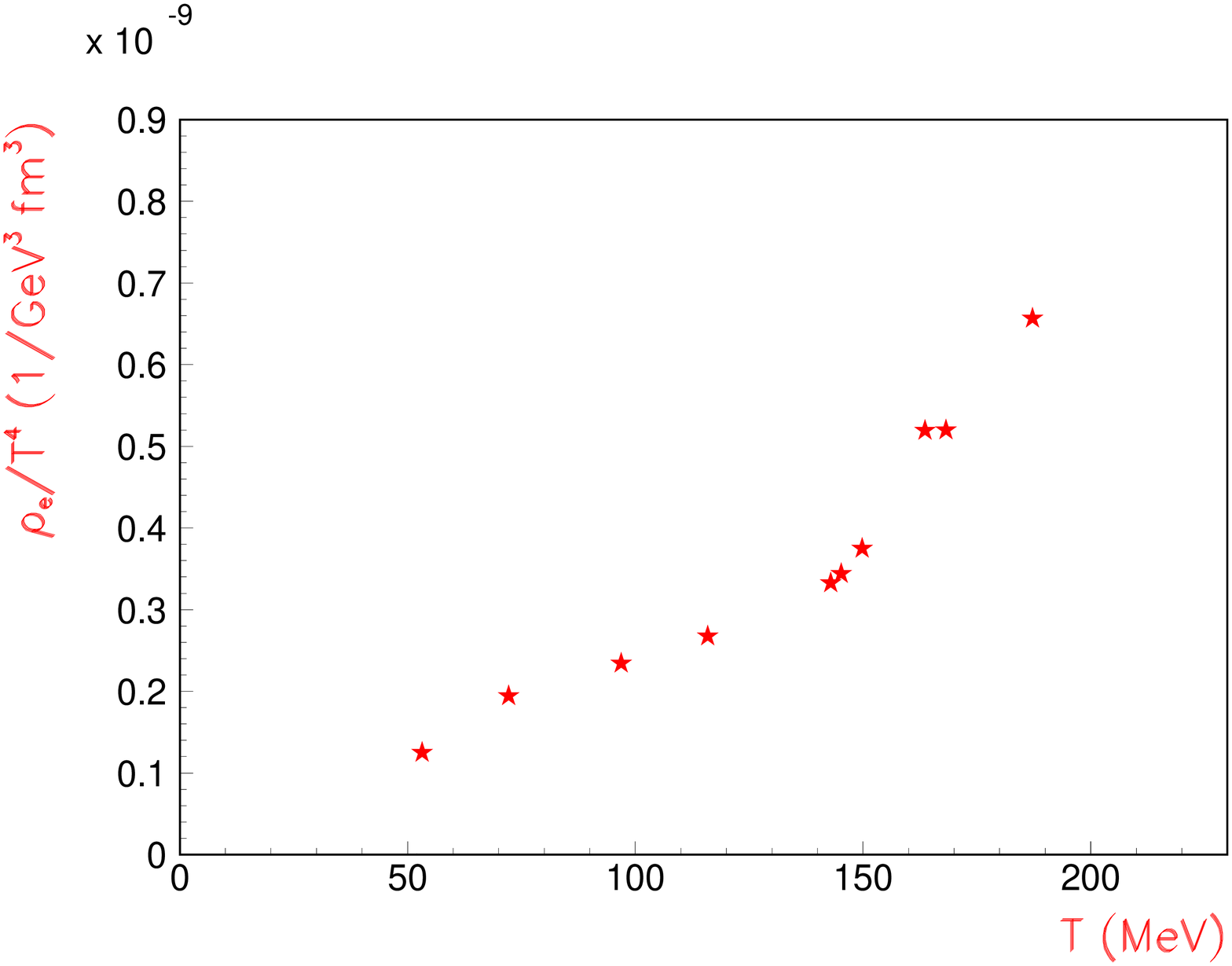,width=120mm}}
\end{center}
\caption{
The ratio of the energy density to the temperature to the
4th power as a function of the temperature.
The temperature is extrapolated to zero fugacities along an isentropic path
for many nucleus+nucleus, hadron+hadron and
lepton+lepton collisions.
We demand for the thermal model fits confidence level $>$ 10\%.
}
\label{rhoe_to_t4_vs_t}
\end{figure}

\begin{figure}[ht]
\begin{center}
\mbox{\epsfig{file=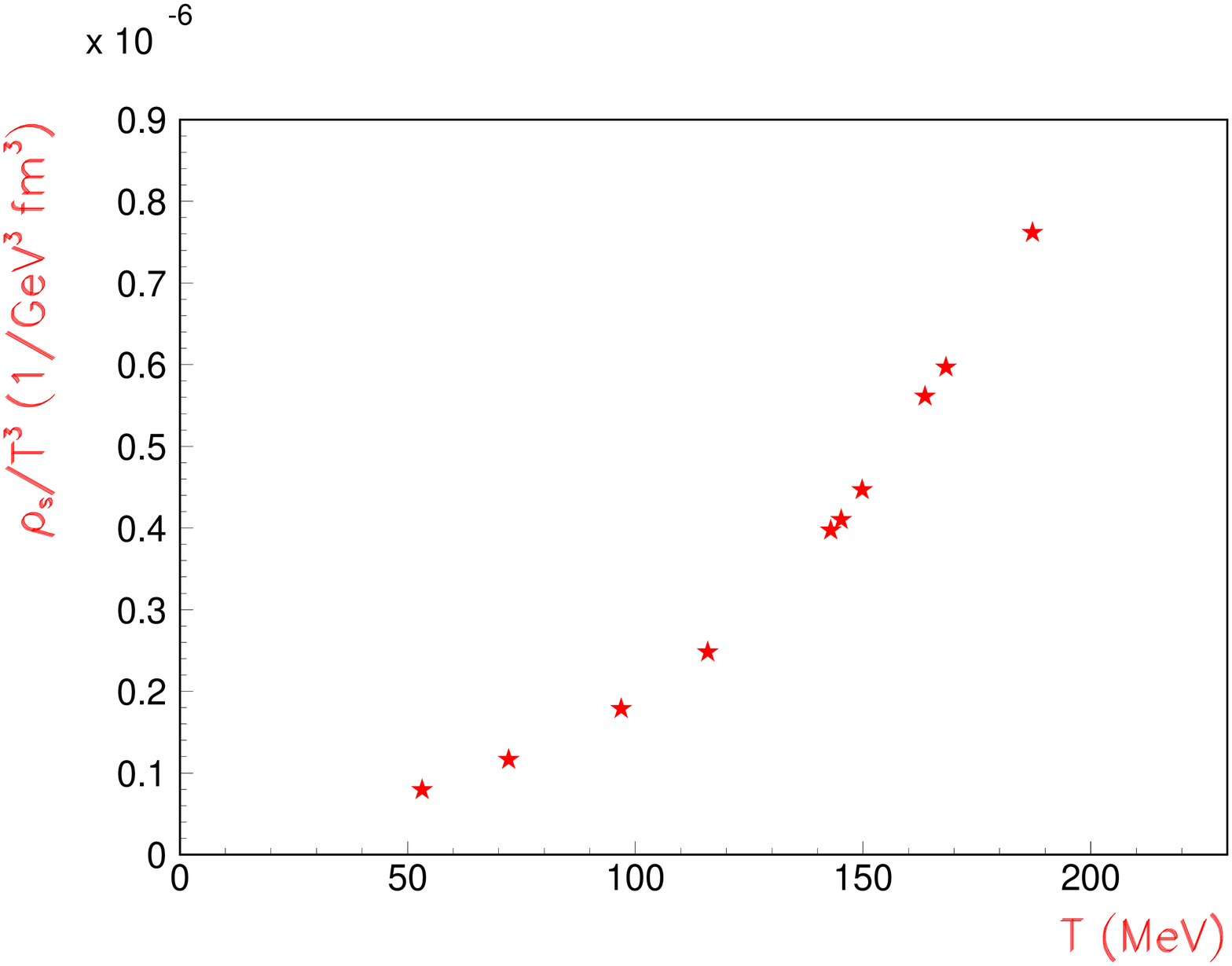,width=120mm}}
\end{center}
\caption{
The ratio of the entropy density to the temperature to the
3th power as a function of the temperature.
The temperature is extrapolated to zero fugacities along an isentropic path
for many nucleus+nucleus, hadron+hadron and
lepton+lepton collisions.
We demand for the thermal model fits confidence level $>$ 10\%.
}
\label{rhos_to_t3_vs_t}
\end{figure}

\begin{figure}[ht]
\begin{center}
\mbox{\epsfig{file=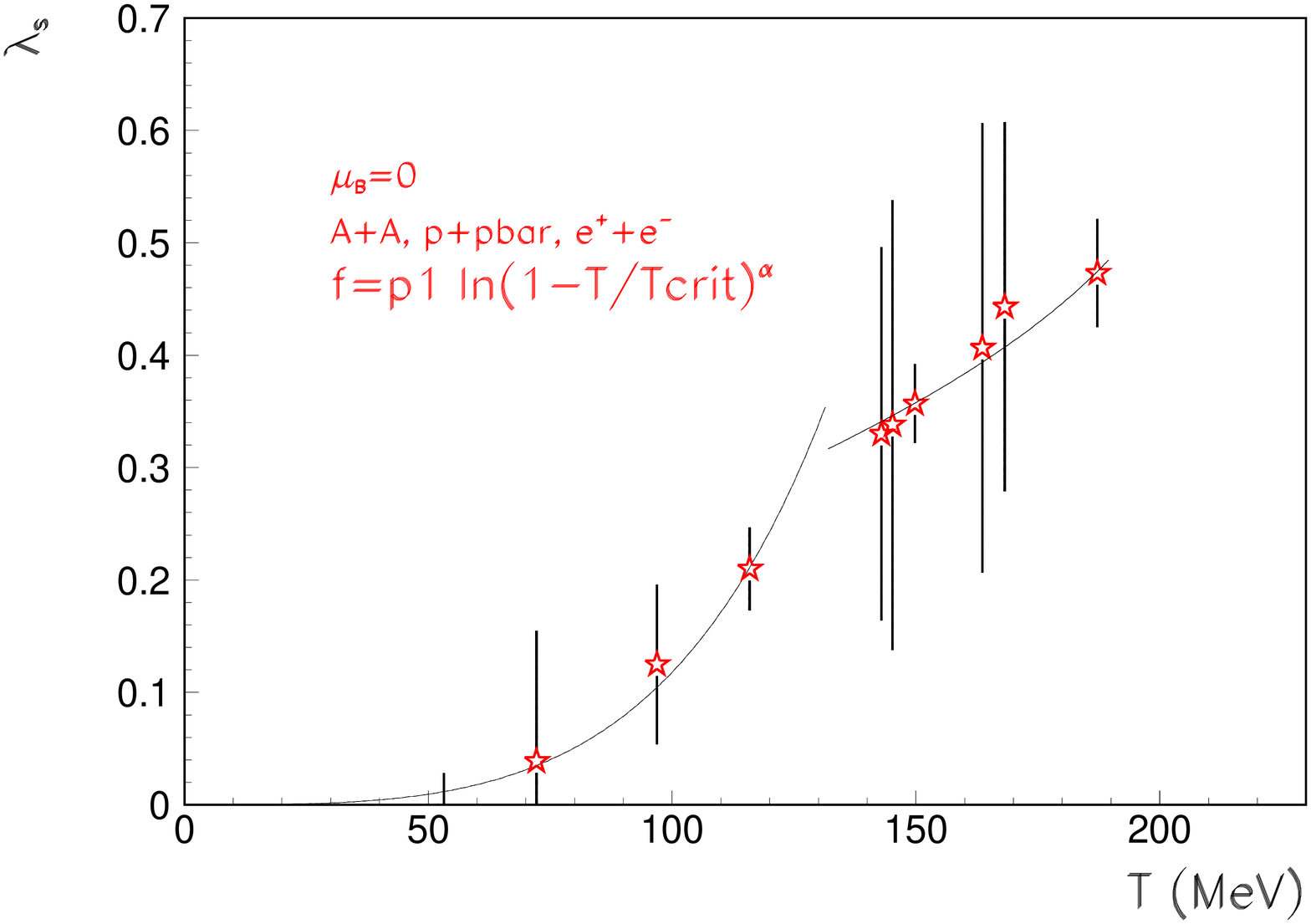,width=120mm}}
\end{center}
\caption{
The $\lambda_s$ factor as a function of the temperature both
extrapolated to zero fugacities along an isentropic path
for many nucleus+nucleus, hadron+hadron and
lepton+lepton collisions.
We demand for the thermal model fits confidence level $>$ 10\%.
The lines shown are fits of the function 
$f= \beta ln (1-T/T_{crit})^{ \alpha} $ below and above
T=140 MeV.
The fit of this function to the data above T=140 MeV
gives $T_{crit}= 218 \pm 70$ MeV, exponent $\alpha=0.54 \pm 0.47$
and $\chi^2/DOF$ = 0.059/3.
}
\label{ls_vs_t}
\end{figure}

\noindent
We study the way the systems approach the
transition in  figures 
\ref{rhoe_vs_t}, \ref{rhos_vs_t},
\ref{rhon_vs_t},
\ref{rhoe_to_rhon_vs_t},
\ref{rhos_to_rhon_vs_t},
\ref{rhos_to_rhoe_vs_t},
\ref{rhoe_to_t4_vs_t},
\ref{rhos_to_t3_vs_t}, \ref{ls_vs_t}.
\\
In figures 
\ref{rhoe_to_rhon_vs_t} and \ref{rhos_to_rhoe_vs_t}
it is seen that the energy density over the particle density
and the entropy density over the energy density
show an onset of a critical behaviour at $T \sim$ 150 MeV,
while the entropy density over the particle density
shown in figure \ref{rhos_to_rhon_vs_t}
saturates above $T \sim$ 150 MeV.
\\
In figure \ref{ls_vs_t}
we try to fit the function
$f= \beta \ [ln(\ 1./ (1 - T/ T_{crit}))]^{ \alpha}$
(logarithmic critical behaviour)
to the data which  go through the 
QGP phase transition according to figures \ref{t} and
\ref{0} (that is the points with $T \geq 145$), 
and extract the critical temperature $T_{crit}^{fit}$ and
the critical exponent $\alpha^{fit}$ from the fit.
We 
find a value for the critical
temperature of 
$T_{crit}^{fit} = 218 \pm 70$ MeV, and an 
exponent $\alpha^{fit}=0.54 \pm 0.47$
with $\chi^2/DOF$ = 0.059/3.
This behaviour can not be studied with the
precision of a study of the neighbourhood of the Curie point in
ferromagnets \cite{ferro}.
\\
This value for $T_{crit}^{fit}$ is 
in agreement with the QCD expectations for the critical
temperature of $T_{crit}^{th}$ = 194 $\pm$ 18 MeV \cite{mapping}.
\\

\section{Conclusions}

\noindent
The starting point of this paper is the extraction
of 
 thermodynamic parameters describing the
 final state of Au+Au collisions at 2 and 4 GeV per nucleon
 and of Pb+Pb collisions at 40 GeV per nucleon.
 We extrapolate these parameters to zero chemical potentials
 along an isentropic path and study the
 strangeness suppression factor $\lambda_s$ ($\lambda_s$=
 $\frac {2 \overline{s} } { \overline{u} + \overline{d} }$)
as a function of the energy density
reached early 
in each collision (initial energy density $\epsilon_i$).
\\

\noindent
We then arrive at the following conclusions:
\\

\noindent
1) The so called 'strangeness suppression' 
puzzle, namely the decrease of the $K/\pi$ ratio
(or equivalently of $\lambda_s$)
with $\sqrt{s}$ increasing from its value in Pb+Pb
collisions at 40 A GeV, is explained
as reflecting the varying chemical potentials
of the heavy ion systems.
\\

\noindent
2) Several other
experimental observations have the same origin:
\\
$\bullet$ the increase of the double ratio $K/\pi$ (A+A/p+p)
with decreasing $\sqrt{s}$.
\\
$\bullet$ the flatter behaviour of the $K/\pi$ ratio
as a function of $\sqrt{s}$ when extracted at midrapidity.
\\
$\bullet$ the difference in the $\sqrt{s}$ dependence
of $K^+/\pi^+$ and $K^-/\pi^-$ ratios.
\\
$\bullet$
the enhancement seen in strange particles in central p+A
collisions as compared to p+p collisions at the same $\sqrt{s}$.
\\

\noindent
3) The $\lambda_s$ value of systems with nonzero baryochemical potential
is found to approach its limiting value at zero $\mu_B$
as defined by the $p \overline{p}$ and $e^+ e^-$ colliding systems.
We estimate that 
the limiting $\lambda_s$  value
(as well as an approximately  net
baryon free midrapidity region) 
will be reached by nuclear collisions
at the initial energy density of $\sim$ 8-9 GeV/$fm^3$, 
 (corresponding
approximately to $\sqrt{s}$ $\sim$ 3-8 TeV per nucleon+nucleon pair)
probably  at the LHC.
\\

\noindent
4) Strangeness is not significantly increased in nucleus nucleus collisions
as compared to elementary particle collisions, if
they are compared 1) at the same (zero) chemical potential
and 2) at the same initial energy density.
\\

\noindent
5) However, $\lambda_s$ is found to significantly increase
in all systems which reach $\epsilon_i$ higher than
$\sim $ 1 GeV/$fm^3$, as compared to all systems
below.
Strangeness is found to follow closely the temperature,
rising until $\epsilon_i$ $\sim $ 1 GeV/$fm^3$ and saturating 
along the border of the QCD phase transition, namely above
1 GeV/$fm^3$.
This allows us to extract in a model independent way
the critical parameters of the QCD phase transition from the data
in particular $\epsilon_{critical}$ = 1 $\pm$ 0.3 GeV/$fm^3$
as well as the limiting T and $\lambda_s$ values \cite{mapping}.
\\

\noindent
6)
Having established in figures \ref{0}, and \ref{t}
the critical initial energy density for the
QCD phase transition of
$\epsilon_{crit}$ $\sim 1 \pm 0.3 GeV/fm^3$,
and determined the systems which go through the QGP phase,
we study the way thermodynamic parameters
approach the transition point.
\\
We find that the systems with $\epsilon_i > 1 GeV/fm^3$,
approach $T_{crit}^{fit}$ = 218 $\pm$ 70 MeV with a
(logarithmic) critical exponent
$\alpha^{fit}$ = 0.54 $\pm$ 0.47.
\\
More data from SPS, RHIC and LHC at several $\sqrt{s}$ will
serve to narrow down the approach to the transition point.

\subsection*{Acknowledgements}

I wish to thank P. Minkowski and K. Pretzl  for stimulating and 
fruitfull discussions and critical comments.
I also wish to thank J. Rafelski, K. Redlich,
H. Oeschler and M. Gorenstein for illuminating discussions.

\newpage

\begin{center}
\begin{figure}[ht]
\vspace*{-0.5cm}
\mbox{\epsfig{file=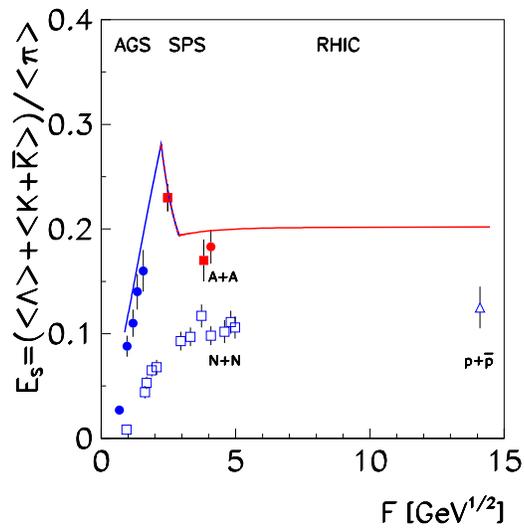,width=120mm}}
\caption{
($ <\Lambda> + <K + \overline{K}>$)$/\pi$ ratio as a function of
F=f($\sqrt{s})$ in A+A and p+p collisions
\protect\cite{na49_kpi}, compared with a model
from reference \protect\cite{marek}.
}
\label{marek}
\vspace*{-0.5cm}
\end{figure}
\vspace*{-0.5cm}
\end{center}

\clearpage

\vspace{1.0cm}


\begin{thebibliography}{99}


\bibitem{lattice} A. Ali Khan et al., CP-PACS collaboration, hep-lat/0008011, \\
	  F. Karsch, E. Laermann, A. Peikert, Ch. Schmidt, S. Stickan, hep-lat/0010027. 


\bibitem{GL} J. Gasser, H. Leutwyler,
 Nucl. Phys. B 307 (1988) 763.
 
\bibitem{PM} P. Minkowski, {\it Czech J. Phys.} {\bf B40} (1990) 1003.


\bibitem{mapping} S. Kabana, P. Minkowski,
hep-ph/0010247, to appear in New J. of Phys.

\bibitem{qmconference} Proceedings of Quark Matter conferences.

\bibitem{stock} R. Stock, Phys. Lett. B 456 (1999) 277.


\bibitem{heinzjakob}
U. Heinz, M. Jacob, nucl-th/0002042.

\bibitem{satz} H. Satz, Rept. Prog. Phys. 63 (2000) 1511.

\bibitem{horst} 
D. Zschiesche et al., 
contribution to the 'Symposium on Fundamental Issues in Elementary
Matter', 25 - 29 September 2000, Bad Honnef, Germany, 
nucl-th/0101047.
\\
S. Scherer et al., Prog. Part. Nucl. Phys. 42 (1999) 279.


\bibitem{rafelski1}  J. Rafelski, GSI Report 81-6 (1981) 282.
\\
J. Rafelski, R. Hagedorn,
Statistical Mechanics of Quarks and Hadrons,
North Holland, Amsterdam, ed. H. Satz (1981).
\\
J. Rafelski, Phys. Rep. 88 (1982) 331.
\\
P. Koch, B. Mueller, J. Rafelski,
Phys. Rep. 142 (1986) 167.


\bibitem{Sonja} 
S. Kabana, 
J. of Phys. G Vol. 27 Nr. 3 (2001) 497, hep-ph/0010228.
\\  
S. Kabana, Proc. of the XXX. Int. Conf. on High Energy Physics, Osaka  2000,
hep-ph/0010246.   


\bibitem{jpsi_prediction}
T. Matsui, H. Satz,
Phys. Lett. B 178 (1986) 416.

\bibitem{horst_charm}
L. Gerland, L. Frakfurt, M. Strikman, H. Stocker, W. Greiner,
J. Phys. G 27 (2001) 695.
\\
M. Gorenstein, A. P. Kostyuk, H. Stocker, W. Greiner, hep-ph/0012015.
\\
P. Braun-Munzinger, J. Stachel, Phys. Lett. B490 (2000) 196.
\\
P. Csizmadia and P. Levai, hep-ph/0008195.
\\
P. Levai et al., hep-ph/0011023.
\\
R. L. Thews, M. Schroedter, J. Rafelski,
hep-ph/0007323.
\\
%
J.P. Blaizot, M. Dinh, J.Y. Ollitrault,
Phys. Rev. Lett. 85 (2000) 4012.
\\
D. Kharzeev, R.L.Thews, Phys. Rev. C 60 (1999) 041901, nucl-th/9907021.
\\
A. Capella, E. Ferreiro, A. Kaidalov,
Phys. Rev. Lett. 85 (2000) 2080.



\bibitem{moriond} 
S. Kabana, talk presented in the 
CERN Heavy Ion Forum, March 13th, 2001.
\\
S. Kabana, to appear in the proceedings of the XXXVIth Rencontres de
Moriond on QCD and high energy hadronic interactions,
17-24 March 2001, Les Arcs 1800, France.


\bibitem{SKPM}
P. Minkowski, W. Ochs, Eur. Phys. J. C9 (1999) 283, hep-ph/9811518.
\\
S. Kabana and P. Minkowski, 
    {\it Phys. Lett.} {\bf B472} (2000) 155, hep-ph/9907570.
    \\
    S. Kabana and P. Minkowski, Proceedings of the
 International Europhysics Conference on High Energy Physics
 (HEP'99), 15-21 July 1999, Tampere, Finland, (IoP publishing),
  page 862, hep-ph/9909351.
  \\
  P. Minkowski, S. Kabana, W. Ochs, Proceedings of the XXX International
  Conference on
  High Energy Physics (ICHEP'2000), 27 July-2 August 2000, Osaka, Japan.
  hep-ph/0011040.

\bibitem{revions1} F. Becattini, J. Cleymans, A. Ker\"{a}nen,
       E. Suhonen and K. Redlich, hep-ph/0002267. 



\bibitem{hagedorn} R. Hagedorn, {\it Nuovo Cim. Suppl.} {\bf 3} (1965) 147.

\bibitem{Gerber} P. Gerber, H. Leutwyler, {\it Nucl. Phys.} {\bf B321} (1989) 387. 

\bibitem{revions} P. Braun-Munzinger and J. Stachel,
   {\it Nucl. Phys.} {\bf A606} (1996) 320-328, nucl-th/9606017.



\bibitem{rafelski} J. Letessier, J. Rafelski, nucl-th/0003014.

\bibitem{biro} T.S. Biro, P. Levai, J. Zimanyi, hep-ph/9807303.

\bibitem{becattee} F. Becattini, A. Giovannini and S. Lupia,
     {\it Z.Phys.} {\bf C72} (1996) 491, hep-ph/9511203.

\bibitem{hepph9702274} F. Becattini and U. Heinz, {\it Z.Phys.} {\bf C76}
(1997) 269, hep-ph/9702274.

\bibitem{rischke}
D. Rischke, to appear in the proceedings of the QM2001
(http://www.rhic.bnl.gov/qm2001/).


\bibitem{pinkenburg}
C. Pinkenburg et al, (E895 coll.),
to appear in the proceedings of the 
QM2001.

\bibitem{e917}
L. Ahle et al., (E866 and E917 coll.), nucl-ex/0008010.
\\
B.B. Back et al., (E917 coll.),
nucl-ex/0003007.
\\
L. Ahle et al., (E866 and E917 coll.),
nucl-ex/9910008.
\\
R. Seto et al., (E917 coll.),
Nucl. Phys. A 638 (1998) 407c-410c.

\bibitem{40gevpbpb}
N. Carrer et al., (NA57 coll.), to appear in the proceedings 
of QM2001.

\bibitem{na49_kpi}
F. Sikler, (NA49 coll.), ISMD 2000, Tihany, Hungary, Oct 2000, hep-ex/0102004.
\\
C. Blume et al., (NA49 coll.),  to appear in the proceedings
of QM2001.

\bibitem{hepph0004138} S. Kabana, hep-ph/0004138.  



\bibitem{enhancement}
F. Antinori et al., (WA97 coll.), Phys. Lett. B 449 (1999) 401.
\\
F. Antinori et al., (WA97 coll.), Phys. Lett. B 433 (1998) 209.
\\
%
C. A. Ogilvie et al., (E802 coll.),
Nucl. Phys. A 630 (1998) 571.
\\
%
L. Ahle et al., (E866 and E917 coll.),
Phys. Lett. B 490 (2000) 53-60.
\\
L. Ahle et al., (E866 and E917 coll.),
Phys. Lett. B 476 (2000) 1-8.
\\
R. A. Barton et al., (NA49 coll.),
J. of Phys. G, Vol. 27 Nr. 3 (2001), 367.
\\
F. Sikler et al., (NA49 coll.), {\it Nucl. Phys.} {\bf A661} (1999) 45.
\\
S. Kabana et al., (NA52 coll.),
J. of Phys. G Vol. 27 Nr 3, (2001) 495, hep-ex/0010053.
\\
S. Kabana et al., (NA52 coll.),  paper submitted to ICHEP2000, hep-ex/0010045.
\\
G. Ambrosini et al. (NA52 coll.), 
   {\it New J. of Phys.} {\bf 1} (1999) 22.
   \\
G. Ambrosini et al. (NA52 coll.), 
   {\it New J. of Phys.} {\bf 1} (1999) 23.
   \\
S. Kabana et al. (NA52 coll.), Nucl. Phys. A 661 (1999) 370c.  
\\
S. Kabana et al. (NA52 coll.), J. of Phys. G, Vol. 25 (1999) 217.
\\
G. Ambrosini et al., (NA52 coll.), Phys. Lett. B 417 (1998) 202.
\\
S. Kabana et al. (NA52 coll.), J. of Phys. G, Vol. 23 (1997) 2135.
\\ 
S. Kabana et al. (NA52 coll.), Nucl. Phys. A 638 (1998) 411c.
\\ 
R.\ Klingenberg et al.,  (NA52 coll.),Nucl. Phys. A 610 (1996) 306c.
\\
I. Bearden et al., (NA44 coll.),
Phys. Lett. B 471 (1999) 6-12.
\\\
D. Roehrich, J. of Phys. G Vol. 27, Nr 3., (2001) 355.
\\
W. Retyk et al., (NA35 coll.), J. of Phys. G (1997) 1845.
\\
T. Alber et al., (NA35 coll.), {\it Z. Phys.} {\bf C64} (1994) 195. 
\\
T. Alber et al., (NA35 coll.), Phys. Lett. B 366 (1996) 56.
\\
J. B\"{a}chler et al., (NA35 coll.), {\it Z. Phys.} {\bf C58} (1993) 367. \\
J. Bartke et al., (NA35 coll.), {\it Z. Phys.} {\bf C48} (1990) 191.


\bibitem{marek}
M. Gorenstein, M. Gazdzicki,
Acta Phys. Pol. 
B 30, (1999) 2705.




\bibitem{inter}
G. Torrieri, J. Rafelski, hep-ph/0103149.
\\
J. Rafelski et al., nucl-th/0101025.
\\
C. Greiner, nucl-th/0012093 and  nucl-th/0009036.
\\
S. Soff et al., J. Phys. G 27 (2001) 449, nucl-th/0010103.
\\
S. Hamieh et al., Phys. Lett. B 486 (2000) 61, hep-ph/0006024.
\\
A. Capella, C.A. Salgado, hep-ph/0007236.


\bibitem{redlichqm2001}
K. Redlich, talk given in QM2001 (http://www.rhic.bnl.gov/qm2001/), 
to appear in the proceedings of the QM2001.

\bibitem{ogilvie_qm2001}
C. A. Ogilvie, to appear in the proceedings of the QM2001.

\bibitem{Sonja2} K. Pretzl et al., (NA52 coll.),
contribution to the 'Symposium on Fundamental Issues in Elementary
Matter', 25 - 29 September 2000, Bad Honnef, Germany, 
nucl-ex/0011016.

\bibitem{myphd}
S. Kabana,
Ph. D. thesis, University of Frankfurt am Main, 1994.

\bibitem{pA}
T. Susa et al., (NA49), to appear in the proceedings of QM2001.
\\
B. Cole, to appear in the proceedings of QM2001.

\bibitem{bjorkenform} J. D. Bjorken, {\it Phys, Rev.} {\bf D27} (1983) 140.


\bibitem{ferro}
Physics of Critical Fluctuations, Y. Ivanchenko and
A. Lisyansky, Springer, (1995).

\end{thebibliography}
\end{document}